\documentclass[twocolumn]{aastex63}

\usepackage{amsmath}
\usepackage{natbib}
\usepackage{multirow} 
\usepackage{anyfontsize}
\usepackage{indentfirst}
\usepackage{comment}
\usepackage{booktabs}
\usepackage{bigints}
\usepackage{mathrsfs,amsmath} 
\usepackage{makecell}
\usepackage{graphicx}

\usepackage[normalem]{ulem}
\useunder{\uline}{\ul}{}
\usepackage{hyperref}
\usepackage{lineno}

\hypersetup{linkcolor=magenta,citecolor=cyan,filecolor=yellow,urlcolor=blue}

\received{ddmmyyyy}
\revised{\today}
\accepted{ddmmyyyy}
\published{ddmmyyyy}

\submitjournal{ApJ}

\shorttitle{Double-trigger Gamma-ray Bursts and Their Classification.}
\shortauthors{Li.}

\begin{document}

\title{Classification and Characteristics of Double-trigger Gamma-ray Bursts}

\author[0000-0002-1343-3089]{Liang Li}

\affiliation{Institute of Fundamental Physics and Quantum Technology, Ningbo University, Ningbo, Zhejiang 315211, People's Republic of China}

\affiliation{School of Physical Science and Technology, Ningbo University, Ningbo, Zhejiang 315211, People's Republic of China}

\correspondingauthor{Liang Li}
\email{liliang@nbu.edu.cn}

\begin{abstract}

Over the past two decades, the \textit{Swift} and \textit{Fermi} missions have identified a rare class of ``double-trigger'' gamma-ray bursts (GRBs) that produce two independent trigger events. These events are characterized by a sufficiently long quiescent period during which the on-board trigger system can reset, resulting in the subsequent emission being recorded as a second independent event. Consistent sky localization confirms that both trigger events originated from the same astrophysical source. Here, we present a systematic classification and characteristics study of three such cases: GRB 091024A, GRB 110709B, and GRB 220627A. We investigate each trigger episode emission independently using standard classification diagnostics, including duration ($T_{90}$), hardness ratio, minimum variability timescale (MVT), spectral lag ($\tau_{\rm lag}$), peak energy ($E_{\rm p}$), and energetics. We compare these properties with those of typical long GRBs (LGRBs) and with single-trigger LGRBs that exhibit extended quiescent periods. Our analysis reveals that all sub-bursts from the three double-trigger events consistently lie within the LGRB classification region. These results indicate that double-trigger GRBs are not a physically distinct subclass, but rather products of LGRB central engines that undergo extended dormancy and subsequent reactivation.

\end{abstract}


\section{Introduction} \label{sec:intro}

Gamma-ray bursts (GRBs) are among the most energetic explosive phenomena in the Universe. The traditional classification scheme divides them into long-soft and short-hard populations, broadly associated with the collapse of massive stars (Type II) and the merger of compact binary objects (Type I), respectively \citep{Kouveliotou1993, Woosley2006, Zhang2009b,Berger2014}.

Observationally, the light curves exhibit diverse temporal morphologies. Some events show a precursor followed by a main emission episode, others exhibit a main pulse with subsequent weaker emission, and still others consist of multiple pulses of comparable intensity. The pulses are often separated by quiescent intervals, though in some cases they exhibit slight temporal overlap. In a subset of multi-pulse GRBs, quiescent periods separate adjacent pulses, intervals during which gamma-ray emission drops to near-background levels, indicating that central-engine activity can be highly intermittent and the complexity of prompt emission light curves points to the intricate nature of central engine activity\citep{Fishman1995,RamirezRuiz2001a,RamirezRuiz2001b,Hakkila2004,Nakar2002,Norris2005,Drago2007}. In particular, the presence of quiescent periods implies that the energy injection into the jet is not always continuous. Most quiescent periods in multi-pulse GRBs are relatively short and occur within single-triggered events \citep[e.g.,][ and references therein]{Li2019a,Li2021b}. A rare subset, however, the quiescent interval exceeds the satellite's background-reset threshold, causing the subsequent emission to be recorded as a second, independent trigger. These events are referred to as double-trigger GRBs \citep{Gruber2011,ZhangBB2012,HuangYY2022}. 

Pulse-resolved analysis reveals that pulses are the fundamental building blocks of GRB light curves. Interestingly, the distribution of interpulse intervals resembles that of pulse widths when quiescent segments are excluded, though quiescent segments themselves show a distinct excess at longer timescales, spanning seconds to hundreds of seconds \citep{Nakar2002}. Multi-episode events with extended quiescent periods thus provide direct probes of the intermittent nature of central-engine activity and jet radiation processes.

The physical origin of quiescent periods remains debated. Two main scenarios have been proposed. First, the central engine may genuinely enter a dormant state during the gap, with both emission episodes powered by essentially the same engine configuration. Second, the engine might remain active throughout, but radiation efficiency or beaming dramatically decreases during certain phases, creating apparent gaps in the observed light curve \citep{Nakar2002, Drago2007}. Statistical studies have found that emission before and after quiescent periods often shows similarities in temporal structure, hardness ratio, and energy release, supporting the ``engine-dormancy'' interpretation for at least some long GRBs \citep{Drago2007}. However, individual case studies also suggest that low-level emission may persist during gaps in some events, complicating the question of whether quiescent periods truly represent complete engine shutdown \citep{Virgili2013, Romano2006}.

GRB classification originally relied on the bimodal distribution of duration and hardness ratio \citep{Kouveliotou1993}. While the classical $T_{90} \sim 2$~s boundary shifts depending on instrument energy bands and trigger conditions, it remains an important empirical benchmark \citep{QinY2013}. As samples have grown and multi-wavelength data have accumulated, additional diagnostics tied more directly to physical origin have been developed. These include spectral lag and its correlation with luminosity \citep{Norris2000, Norris2006}, minimum variability timescale \citep{Golkhou2014}, and empirical correlations between spectral peak energy and energetics such as the Amati and Yonetoku relations \citep{Amati2002, Yonetoku2004}. 
Additionally, composite parameters like the $\varepsilon$ parameter (constructed from $E_{\gamma,\rm iso}$ and $E_{p,z}$) and three-dimensional classification frameworks incorporating the amplitude parameter $f$ provide more robust discrimination in ambiguous  cases, such as short GRBs with extended emission or pseudo-short GRBs at high redshift \citep{Lv2010, Lv2014}. These diagnostics enable physical classification of individual events and their constituent sub-bursts.

Following these lines of argument, well-studied double-trigger events such as GRB~091024A \citep{Gruber2011b}, GRB~110709B \citep{ZhangBB2012}, and the more recent GRB~220627A \citep{deWet2023} offer unique insights into central-engine timescales. The key question is whether these double-trigger events represent a distinct GRB subclass. Specifically, does the mechanism producing exceptionally long quiescent periods reflect unusual central-engine structures—such as two-stage collapses or magnetar reactivation following a binary neutron star merger?Or is double-triggering simply the extreme tail of the intermittency distribution characteristic of normal long GRB central engines?

To address this question, we analyze confirmed double-trigger GRBs, treating each triggered sub-burst as an independent event. We apply standard classification diagnostics and compare the results with two control samples: normal long GRBs and single-trigger long GRBs with extended quiescent periods. In Section~\ref{sec:Methodology}, we introduce our sample selection and data reduction. Section~\ref{sec:results} presents our classification results and comparative analysis. In Section~\ref{sec:discussion}, we discuss physical models and their implications for a unified picture of GRB central engines. We summarize our conclusions in Section~\ref{sec:conclusions}.

\section{Methodology}\label{sec:Methodology}

We focus on three double-trigger GRBs observed by \textit{Swift}/BAT and \textit{Fermi}/GBM: GRB~091024A, GRB~110709B, and GRB~220627A. These events satisfy three criteria: (1)~two independent triggers were recorded from consistent sky positions, confirming they originated from the same astrophysical source; (2)~the quiescent period between triggers was long enough (several hundred seconds) for the trigger system to reset; and (3)~sufficient spectral and temporal data are available for detailed analysis. For each triggered sub-burst, we measured the following parameters.

\subsection{Measured Parameters for individual pulses/episodes.}

\noindent{\bf Pulse-resolved duration $T^{\rm pulse}_{90}$.} For each sub-burst episode (or pulse), we measure the duration $T^{\rm pulse}_{90}$ from the background-subtracted cumulative-counts curve $C(t)$ within a time window $[T_{\rm start},\,T_{\rm stop}]$ that brackets the emission episode. We first identify the maximum cumulative counts in the window, $C_{\max}=\max\{C(t)\}$, and define the 5\% and 95\% levels as $C_{05}=0.05\,C_{\max}$ and $C_{95}=0.95\,C_{\max}$. The corresponding times $T_{05}$ and $T_{95}$ are then obtained by searching the discrete time series for the samples that minimize $|C(t)-C_{05}|$ and $|C(t)-C_{95}|$, respectively. The pulse duration is defined as
\begin{equation}
T^{\rm pulse}_{90}=T_{95}-T_{05}.
\end{equation}
To estimate asymmetric uncertainties, we construct upper and lower envelopes of the cumulative counts using $C^{\rm up}(t)=C(t)+\sigma_C(t)$ and $C^{\rm low}(t)=C(t)-\sigma_C(t)$, and repeat the procedure using $C_{\max}^{\rm up}=\max\{C^{\rm up}(t)\}$ and $C_{\max}^{\rm low}=\max\{C^{\rm low}(t)\}$ to obtain $T_{05}^{\rm up}$, $T_{05}^{\rm low}$, $T_{95}^{\rm up}$, and $T_{95}^{\rm low}$. We define
\begin{equation}
T_{05}^{+}=T_{05}^{\rm up}-T_{05},\qquad T_{05}^{-}=T_{05}-T_{05}^{\rm low},
\end{equation}
\begin{equation}
T_{95}^{+}=T_{95}^{\rm up}-T_{95},\qquad T_{95}^{-}=T_{95}-T_{95}^{\rm low},
\end{equation}
and propagate these to $T_{90}$ by quadrature,
\begin{equation}
\sigma_{T_{90}}^{+}=\sqrt{(T_{05}^{+})^{2}+(T_{95}^{+})^{2}},\qquad
\sigma_{T_{90}}^{-}=\sqrt{(T_{05}^{-})^{2}+(T_{95}^{-})^{2}}.
\end{equation}
We apply this procedure independently to each sub-burst or pulse (e.g., $G_1$ and $G_2$) using its own time window and cumulative-counts file, yielding episode-resolved $T_{90}$ values and uncertainties for subsequent classification and comparisons.

\noindent{\bf Hardness ratio.}
For each identified sub-burst episode, we compute a time-integrated hardness ratio (HR) using background-subtracted light curves extracted in two energy bands. For \emph{Swift}/BAT data, we extract mask-weighted light-curve in a ``low'' band (e.g., 25--50~keV) and a ``high'' band (e.g., 50--100~keV) from the standard BAT \texttt{.lc} FITS files, adopting a uniform time bin size $\Delta t$ (e.g., 256~ms for long GRBs and 64~ms for short GRBs). The trigger time $t_{\rm trig}$ is obtained from the UK Swift Science Data Centre (UKSSDC) repository\footnote{\url{https://www.swift.ac.uk/.}}. We developed an python code called \texttt{my\_pulses.py} to authoritatively identity the number of independent pulses and the episode boundaries $[t_1,t_2]$ (relative to $t_{\rm trig}$) for each GRB. We shift the light-curve time stamps to $t={\rm TIME}-t_{\rm trig}$ and select data within the interval $t\in[t_1,t_2]$ for both energy bands.

Within each band, we convert the binned count rate ${\rm RATE}(t_i)$ to cumulative counts by direct time integration,
\begin{equation}
C(t_j)=\sum_{i\le j} {\rm RATE}(t_i)\,\Delta t ,
\end{equation}
The cumulative-count uncertainty is propagated by adding the per-bin rate errors ${\rm ERROR}(t_i)$ in quadrature,
\begin{equation}
\sigma_C(t_j)=\left[\sum_{i\le j} {\rm ERROR}(t_i)^2\right]^{1/2}\Delta t ,
\end{equation}
treating the uncertainties in different bins as independent. The net accumulated counts in the episode are then evaluated as the difference between the cumulative counts at the episode end and start,
\begin{equation}
\Delta C = C(t_2)-C(t_1),
\qquad
\sigma_{\Delta C}=\left[\sigma_C(t_2)^2+\sigma_C(t_1)^2\right]^{1/2},
\end{equation}
where $t_1$ and $t_2$ correspond to the nearest sampled time bins. The hardness ratio is defined as
\begin{equation}
{\rm HR}=\frac{\Delta C_{\rm high}}{\Delta C_{\rm low}},
\end{equation}
with uncertainty
\begin{equation}
\sigma_{\rm HR}
={\rm HR}\left[
\left(\frac{\sigma_{\Delta C_{\rm high}}}{\Delta C_{\rm high}}\right)^2
+
\left(\frac{\sigma_{\Delta C_{\rm low}}}{\Delta C_{\rm low}}\right)^2
\right]^{1/2}.
\end{equation}
For the \emph{Fermi}/GBM omparison, we adopt the catalog definition and use the fluence ratio between  50--300~keV and 10--50~keV \citep{Goldstein2017}.

\noindent \textbf{Minimum variability timescale.}
We estimate the minimum variability timescale (MVT), $\Delta t_{\min}$, for each emission episode using the Haar-wavelet structure-function method. Following \cite{Golkhou2014,Golkhou2015}, we construct background-subtracted light curves and compute the Haar scaleogram $\sigma_{X,\Delta t}$ as a function of timescale $\Delta t$ using the measured count-rate uncertainties. For temporally smooth variations, the first-order structure function increases approximately linearly with lag, ${\rm SF}(\tau)\propto \tau$, corresponding to the linear-rise phase of the scaleogram at small $\Delta t$. The transition to uncorrelated variability produces a break where $\sigma_{X,\Delta t}$ departs from the linear trend and flattens. We define $\Delta t_{\min}$ as this break timescale, typically associated with the rise time of the narrowest resolved pulse. In practice, we evaluate $\sigma_{X,\Delta t}$ over timescales spanning the native light-curve resolution up to the duration of the selected interval and fit a broken power-law model to the scaleogram, enforcing an initial $\sigma_{X,\Delta t}\propto \Delta t$ behavior followed by a plateau above $\Delta t_{\min}$. This yields $\Delta t_{\min}$ and its $1\sigma$ uncertainty. When the lower bound on $\Delta t_{\min}$ falls below the minimum measurable timescale $\Delta t_{\rm S/N}$ (the first statistically significant scale in the scaleogram), we report an upper limit. For the Swift/BAT analysis, we adopt a signal-to-noise threshold ${\rm S/N}=3$ and apply bootstrap resampling ($n_{\rm repl}=2$). We apply the method to the full prompt interval and, when appropriate, to individual pulse intervals $[t_1,t_2]$ defined by our pulse catalog. Further details of the MVT method are provided by \citet{Golkhou2014} and \citet{Golkhou2015}.

\noindent{\bf Spectral lag $\tau_{\rm lag}$.}
For each burst, we measured the spectral lag $\tau_{\rm lag}$ between two energy-resolved light curves on a pulse-by-pulse basis. The start and stop times of each pulse, $(t_1,t_2)$, are taken from our pulse catalog, with all times referenced to the trigger time $T_0$.
For \emph{Swift}/BAT, we extracted mask-weighted light curves in a low-energy band (e.g., 25--50~keV) and a high-energy band (e.g., 50--100~keV) from the standard \texttt{.lc} products and select the time bins within $[T_0+t_1,\,T_0+t_2]$.
For \emph{Fermi}/GBM, we construct two count light curves by selecting events in two PHA channel ranges and binning them uniformly over the pulse interval.
In both cases, we used the tabulated uncertainties in each time bin to propagate statistical errors.

We quantified the lag using the cross-correlation function (CCF). Given two discrete time series, $f_1(t_i)$ and $f_2(t_i)$, we first remove the mean and normalize each series by its standard deviation, then compute the discrete cross-correlation
\begin{equation}
{\rm CCF}(\tau) \equiv \sum_i \tilde{f}_1(t_i)\,\tilde{f}_2(t_i+\tau),
\end{equation}
where $\tilde{f}_k$ denotes the zero-mean, unit-variance series. We restricted the lag search to a finite window $|\tau|\le \tau_{\rm max}$ to avoid spurious edge effects and adopt the lag $\tau_{\rm lag}$ that maximizes $|{\rm CCF}(\tau)|$. To achieve sub-bin precision, we fit a parabola to the CCF maximum and its two adjacent points and take the vertex as $\tau_{\rm lag}$.

We estimate the lag uncertainty via Monte Carlo bootstrap resampling that accounts for bin-by-bin measurement errors. For each pulse, we generate $N_{\rm boot}=2000$ realizations by perturbing each bin with Gaussian noise whose standard deviation equals the reported error. To reduce discretization effects, we interpolate each realization onto a finer time grid (a resolution increase by a factor of 10) before recomputing the CCF. The lag uncertainty is taken as the standard deviation of the bootstrap distribution, and we report the mean lag with its $1\sigma$ uncertainty.

\noindent\textbf{Spectral analysis.} We perform prompt-emission spectral fitting using a forward-folding likelihood approach implemented in the \texttt{3ML} framework \citep{Vianello2015}. For each burst (or pulse interval, when applicable), we extracted time-integrated spectra from the selected detectors, defined background intervals adjacent to the source window, and construct the corresponding instrument response matrices. We model the prompt emission spectrum with the Band function \citep{Band1993},
\begin{equation}
N_{\rm Band}(E)=A
\begin{cases}
\left(\dfrac{E}{E_{\rm piv}}\right)^{\alpha}\exp\!\left(-\dfrac{E}{E_{0}}\right), & E<(\alpha-\beta)E_{0},\\[6pt]
\left[\dfrac{(\alpha-\beta)E_{0}}{E_{\rm piv}}\right]^{\alpha-\beta}\exp(\beta-\alpha)\left(\dfrac{E}{E_{\rm piv}}\right)^{\beta}, & E\ge(\alpha-\beta)E_{0},
\end{cases}
\end{equation}
where $A$ is the normalization, $E_{\rm piv}$ is fixed to 100~keV, and $(\alpha,\beta)$ are the low-energy and high-energy photon indices. The spectral peak energy in the $\nu F_{\nu}$ representation is
\begin{equation}
E_{\rm p}=(2+\alpha)E_{0}\qquad (\beta<-2),
\end{equation}
with uncertainty propagated from the fitted parameter covariance. We verify the robustness of the best-fit parameters by exploring the likelihood surface around the optimal solution.

\noindent\textbf{Energetics.} 
Using the best-fit spectral parameters, we compute the energy flux by integrating the model spectrum over the detector bandpass and applly a $k$-correction to a common cosmological rest-frame band (1--$10^{4}$~keV). Specifically, the observed energy flux in $[e_1,e_2]$ is
\begin{equation}
F^{\rm obs}_{[e_1,e_2]}=\int_{e_1}^{e_2} E\,N(E)\,{\rm d}E,
\end{equation}
and the $k$-correction factor is defined as
\begin{equation}
k_c=\frac{\int_{E_1/(1+z)}^{E_2/(1+z)} E\,N(E)\,{\rm d}E}{\int_{e_1}^{e_2} E\,N(E)\,{\rm d}E},
\end{equation}
with $(E_1,E_2)=(1,10^{4})$~keV. The isotropic-equivalent peak luminosity $L_{\gamma, \rm p, iso}$ in the rest-frame is
\begin{equation}
L_{\gamma, \rm p, iso}=4\pi D_L^2(z)\,F^{\rm obs}_{[e_1,e_2]}\,k_c,
\end{equation}
and the isotropic-equivalent energy $E_{\gamma,\rm iso}$ is computed from the fluence $S_{\gamma}$ as
\begin{equation}
E_{\gamma, \rm iso}=\frac{4\pi D_L^2(z)\,S_{\gamma}\,k_c}{1+z},
\end{equation}
where $D_L(z)$ is the luminosity distance. These procedures ensure that $E_{\rm p,z}=E_{\rm p}(1+z)$ and the energetics are consistently derived in a uniform rest-frame band.

\subsection{Classification Diagnostics}
We employed multiple classification planes to determine the physical type of each sub-burst.

\subsubsection{~$T_{90}$--hardness ratio plane}
To classify individual emission episodes on a common phenomenological plane, we place each pulse on the standard hardness--duration diagram \citep{Kouveliotou1993,Goldstein2017}. For each pulse, we measure $T_{90}$ from the background-subtracted cumulative counts, defining $T_{90}=T_{95}-T_{05}$, where $T_{05}$ and $T_{95}$ are the times at which 5\% and 95\% of the total accumulated counts within the pulse interval are reached. We then compare the pulse-wise measurements against the empirical distributions of Type~I (short, hard) and Type~II (long, soft) populations, using the canonical reference boundary at $T_{90}=2$~s in the observer frame as a reference. 

To quantify where each pulse lies relative to the two populations in the $\log T_{90}$--$\log HR$ plane, we visualize the short- and long-burst distributions using two-dimensional probability densities and with $1\sigma$ and $2\sigma$ confidence contours from bivariate normal approximations. Treating pulses as independent data points enables a direct test of whether distinct episodes within the same GRB occupy different regions of the hardness--duration space, which would indicate intra-burst changes in spectral hardness and characteristic timescale beyond a single time-integrated classification.

\subsubsection{Minimum variability timescale and the $\Delta t_{\rm min}$--$T_{90}$ diagram}
The minimum variability timescale (MVT), $\Delta t_{\rm min}$, is defined as the shortest observable timescale over which a GRB exhibits statistically significant uncorrelated flux variations. We measure $\Delta t_{\rm min}$ for each identified pulse by applying the Haar-wavelet structure-function method to the background-subtracted light curve \citep{Golkhou2014,Golkhou2015}. In this approach, the flux-variation scaleogram $\sigma_{X,\Delta t}$ is computed as a function of the timescale $\Delta t$ after subtracting the Poisson noise floor. For sufficiently high signal-to-noise ratio, $\sigma_{X,\Delta t}$ exhibits a linear-rise regime at small $\Delta t$ (corresponding to smooth variability) followed by a break to a flatter regime that marks the onset of uncorrelated variability. We estimate $\Delta t_{\rm min}$ by fitting a broken power law to the scaleogram, assuming an initial linear rise with $\Delta t$ and a plateau beyond the break \citep{Golkhou2015}. The uncertainty is obtained by propagating the measurement errors. When the lower bound on $\Delta t_{\rm min}$ falls below the minimum measurable timescale $\Delta t_{\rm S/N}$ set by Poisson statistics, we report an upper limit \citep{Golkhou2015}.

We then use the $\Delta t_{\rm min}$--$T_{90}$ plane as a pulse-wise classification diagnostic. For each pulse, we pair its $T_{90}$ with the corresponding $\Delta t_{\rm min}$ and compare the location to reference distributions of short- and long GRBs derived from large samples \citep{Golkhou2015}. In the \citep{Golkhou2015} Fermi/GBM sample, $\Delta t_{\rm min}$ is bounded from above by $T_{90}$ and shows only marginal correlation with $T_{90}$, making the diagram sensitive to differences in temporal structure rather than duration alone. The median observer-frame $\Delta t_{\rm min}$ values are $\sim 134$ ms for long-duration GRBs and $\sim 18$ ms for short-duration GRBs \citep{Golkhou2015}. Adopting $T_{90}=2$ s as the conventional division, we classify each pulse by its consistency with the short- and long-burst distributions in the $\Delta t_{\rm min}$--$T_{90}$ plane.

\subsubsection{~$E_{\rm p,z}$--$E_{\gamma,\rm iso}$ plane (Amati relation)}
We examine whether individual pulses follow the Amati relation by comparing $E_{\rm p,z}$ with the isotropic-equivalent radiated energy $E_{\gamma,\rm iso}$ in a common rest-frame band \citep{Amati2002,Amati2006}. For the comparison samples of long and short GRBs, we perform power-law regression of the form
\begin{equation}
E_{\rm p,z} = A \left(\frac{E_{\gamma,\rm iso}}{E_{\rm piv}}\right)^{B},
\end{equation}
where $E_{\rm piv}$ is a pivot energy chosen to reduce covariance between $A$ and $B$. 
The best-fit parameters and confidence regions are obtained by fitting the power-law model and propagating uncertainties through the confidence intervals. We then evaluate the placement of the pulse-wise points relative to the best-fit long-GRB locus and its intrinsic scatter, which provides a consistency check for sub-burst classification in double-trigger events.

\subsubsection{~$L_{\gamma,\rm p,iso}$--$\tau$ plane (Norris relation)}
We measured the spectral lag, $\tau$, for each pulse by applying a cross-correlation analysis to two background-subtracted count light curves extracted in a low-energy band and a high-energy band. For \emph{Swift}/BAT, we used the 25--50 and 50--100~keV bands, while for \emph{Fermi}/GBM we adopted channel selections that approximate the original energy separation used in the lag literature (e.g., 25--50 and 100--300~keV) when available. For each pulse, the analysis is restricted to the time interval $[t_1,t_2]$ determined from our pulse selection, with light curves binned to a fixed temporal resolution. The lag is defined as the time shift that maximizes the CCF, with the convention that $\tau>0$ indicates low-energy emission lagging high-energy emission.

To estimate the uncertainty, we generate Monte Carlo bootstrap realizations by perturbing each time bin according to its statistical error, repeat the CCF calculation, and construct the lag distribution from the ensemble. We adopt the distribution mean as the best-fit lag and its standard deviation as the $1\sigma$  uncertainty. We then compare our measurements with the empirical Norris lag--luminosity relation for long GRBs \citep{Norris2000},
\begin{equation}
L_{53}=1.3\left(\frac{\tau}{0.01~{\rm s}}\right)^{-1.14},
\end{equation}
where $L_{53}\equiv L_{\rm p,iso}/10^{53}~{\rm erg~s^{-1}}$. For bursts with known redshift, $L_{\rm p,iso}$ is obtained from the time-resolved Band spectral fit with a $k$-correction to a common rest-frame band (1--$10^{4}$~keV). In the $\tau$--$L_{\rm p,iso}$ plane, consistency with the Norris relation provides an additional diagnostic that is characteristic of Type~II (long-GRB-like) emission, complementing the duration- and hardness-based classifiers.

\subsubsection{~$\varepsilon$ parameter}
We adopt the phenomenological parameter $\epsilon$, defined as
\begin{equation}
\epsilon \equiv \frac{E_{\gamma,\rm iso,52}}{E_{\rm p,z,2}^{5/3}},
\end{equation}
where $E_{\gamma,\rm iso,52}\equiv E_{\gamma,\rm iso}/10^{52}\ {\rm erg}$ and $E_{\rm p,z,2}\equiv E_{\rm p,z}/10^{2}\ {\rm keV}$, as proposed by \citet{Lv2010} to separate Type~I (compact-object origin) and Type~II (massive-star origin) GRBs. For each pulse (or sub-burst), we convert the observed spectral peak energy to the rest frame via $E_{\rm p,z}=E_{\rm p}(1+z)$ and compute $\epsilon$ using the corresponding isotropic-equivalent radiated energy $E_{\gamma,\rm iso}$. The uncertainties on $\epsilon$ are propagated from the measurement errors of $E_{\rm p}$ and $E_{\gamma,\rm iso}$ using standard error propagation implemented with the \texttt{uncertainties} package. Because the errors are often asymmetric, we evaluate the positive and negative contributions separately, yielding asymmetric uncertainties for both $\epsilon$ and $\log_{10}\epsilon$. We place each pulse on the $\log_{10}\epsilon$--$\log_{10}T_{90,z}$ plane, with $T_{90,z}=T_{90}/(1+z)$, and compare with the empirical separation at $\epsilon\simeq 0.03$, where high-$\epsilon$ events ($\epsilon>0.03$) preferentially occupy the Type~I region and low-$\epsilon$ events ($\epsilon<0.03$) populate the Type~II region \citep{Lv2010}.

\subsubsection{Energy-hardness-duration (EHD) parameter}
Following \citet{Minaev2020}, we adopt the dimensionless parameter ${\rm EHD}$ to characterise individual emission episodes. For each episode, we compute the rest-frame duration $T_{90,i}=T_{90}/(1+z)$ and the rest-frame spectral peak energy $E_{\rm p,z}=E_{\rm p}(1+z)$. We further define the scaled quantities $E_{\rm p,z,2}\equiv E_{\rm p,z}/(100~{\rm keV})$ and $E_{\rm iso,51}\equiv E_{\gamma,\rm iso}/(10^{51}~{\rm erg})$, where $E_{\gamma,\rm iso}$ is the bolometric isotropic-equivalent energy. The EHD parameter is then calculated as
\begin{equation}
{\rm EHD} \equiv E_{\rm p,z,2},E_{\rm iso,51}^{-0.4},T_{90,i}^{-0.5}.
\end{equation}
The uncertainties are obtained by standard error propagation applied to $(E_{\rm p},E_{\gamma,\rm iso},T_{90})$. To account for asymmetric measurement errors, we propagate the upper and lower errors separately, yielding asymmetric uncertainties for ${\rm EHD}$.

\noindent
The EHD distribution provides an empirical separator between the two canonical GRB populations \citep{Minaev2020}. Type~I events tend to occupy the high-${\rm EHD}$ region, while Type~II events cluster at low ${\rm EHD}$, with a suggested dividing value around ${\rm EHD}\simeq 2.6$ \citep{Minaev2020}. We place each pulse on the ${\rm EHD}$--$T_{90,i}$ plane and compare its location with reference populations to assess whether different emission episodes of the same GRB are consistent with a common classification or show discordant classifications indicative of distinct physical origins.

\section{Results} \label{sec:results}

We measure each triggered sub-burst independently for all three double-trigger events and project them onto multiple classification planes for comparative analysis. Type~I/II boundary distributions were taken from large \textit{Swift} and \textit{Fermi} GRB samples.

\subsection{GRB 091024A}

\noindent{\bf Observations}.
GRB~091024A triggered the Gamma-ray Burst Monitor (GBM; \citealt{Meegan2009}) on board {\it Fermi} twice. \emph{Fermi}/GBM triggered and localized the first episode at 08:55:58.472 UT on 2009 October 24 (GRB 091024372; ; hereafter $T_{0}$) and triggered again at 09:06:29.357 UT (GRB 091024380), with a trigger time separation of $\Delta t=630.885$ s ($\sim$10.5 minutes)  \citep{2009GCN..9866....1B,Gruber2011,Gruber2011,Virgili2013}. The event also triggered the Burst Alert Telescope (BAT),\emph{Konus-Wind} (KW), and several other other high-energy instruments \citep{2009GCN.10062....1M,2009GCN.10083....1G}. Long-lasting follow-up multiwavelenghth afterglow observations were carried out by the  \textit{Swift} X-ray Telescope (XRT) and several ground-based optical telescopes \citep{2009GCN.10069....1P,2009GCN.10063....1M,2009GCN.10064....1N,2009GCN.10067....1L}.

\begin{figure}[ht!]
\centering
{\bf a}\includegraphics[width=1.0\textwidth]{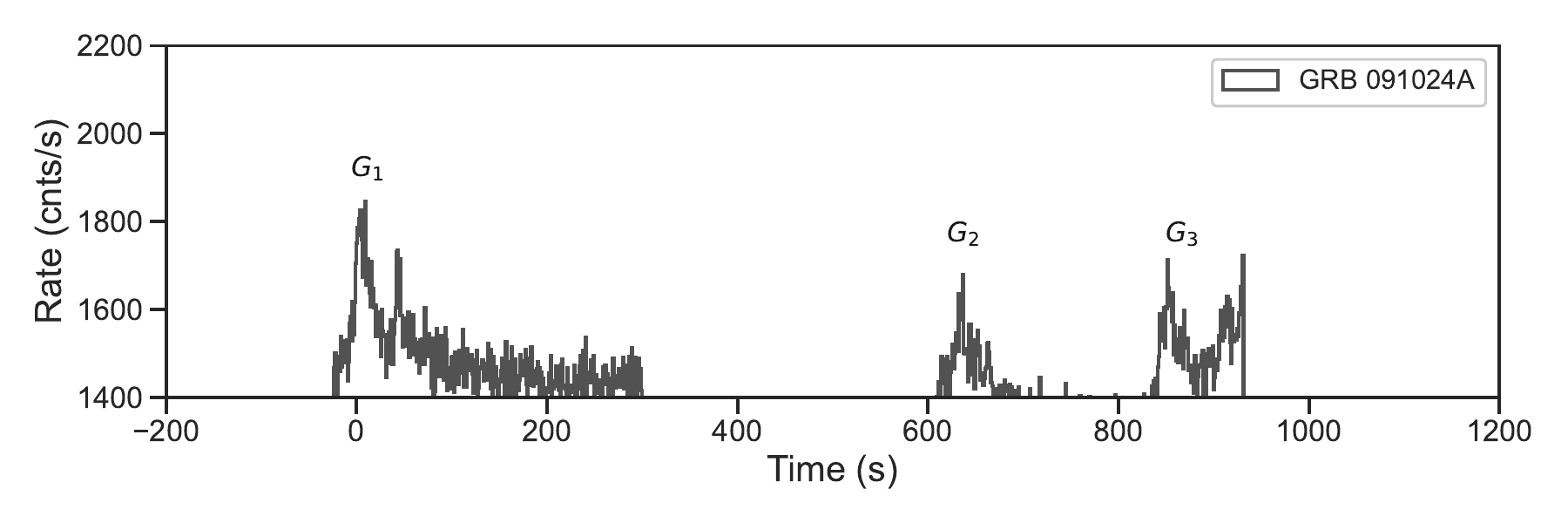}
{\bf b}\includegraphics[width=0.9\textwidth]{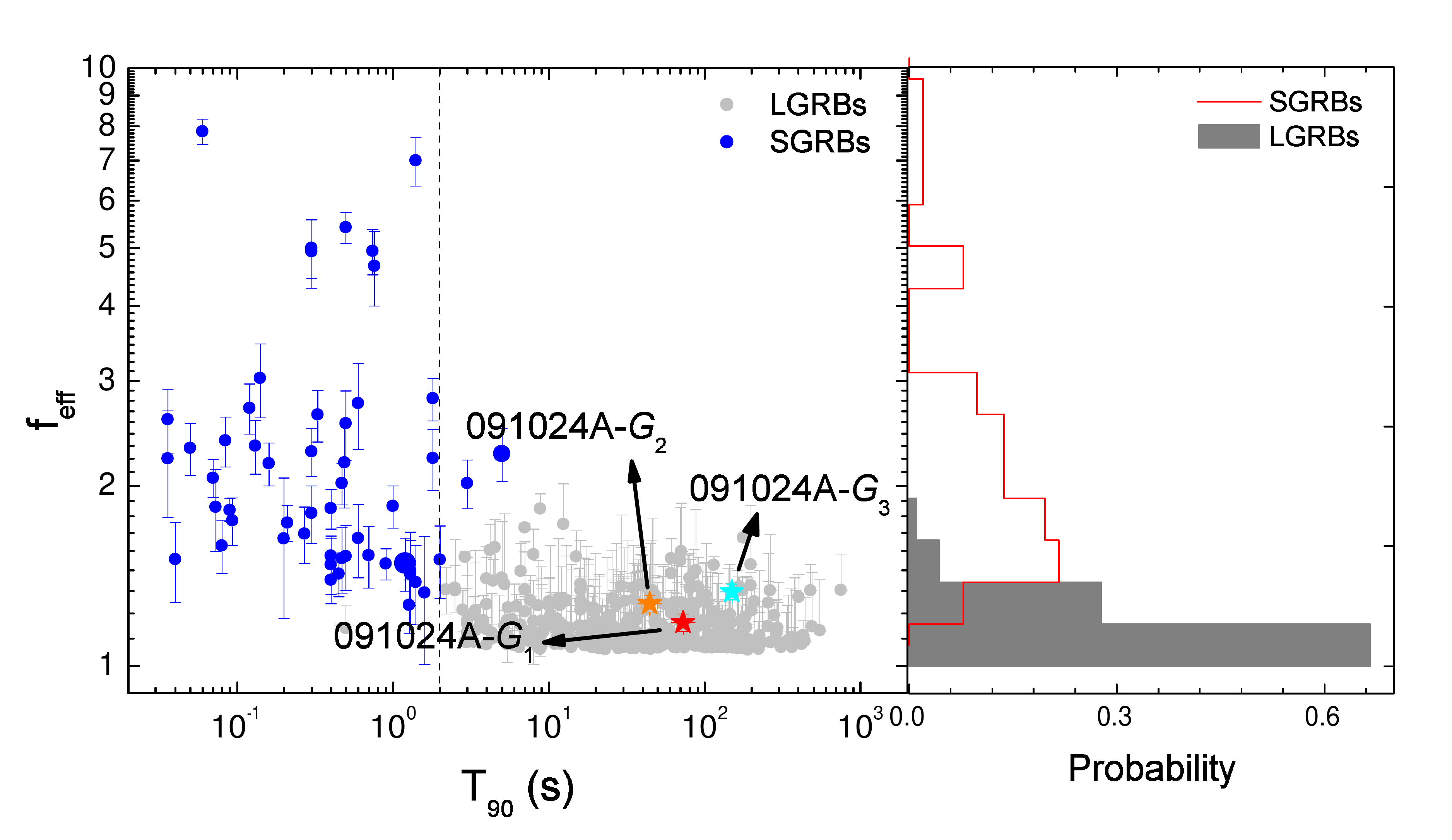}
\caption{{\bf Prompt $\gamma$-ray emission phase of GRB 091024A.} {\bf a}, The background-subtracted light curves of GRB 091024A in distinct energy bands (15-25 keV: green, 25-50 keV: magnetic, 50-100 keV: cyan, 100-350 keV: yellow, 15-350 keV: black), derived from \emph{Swift}/BAT data. {\bf b,}  $T_{90}$-$f(f_{\rm eff})$ plot, where $f$ represents the ratio between the peak flux and the average background flux of a GRB (Methods). $f_{\rm eff}$ is the effective $f$ parameter, indicating how a long GRB can be disguised as a short GRB by arbitrarily lowering its flux level. The three distinct bursts of GRB 091024A are highlighted by the red ($G_1$), orange ($G_2$) and cyan ($G_3$) solid stars, respectively. \label{fig:091024_lc}}
\end{figure}

\noindent{\bf Pulse-resolved or Sub-burst Classification and Characteristics}.
GRB~091024A was detected as an unusually long, multi-episode burst with a total duration of approximately 1020 s. The GBM and KW datasets provide complete temporal coverage. The light curve reveals three distinct multi-peak emission episodes (\citealt{Gruber2011b}, Figure~\ref{fig:091024_lc}). For each episode (denoted $G_1$, $G_2$, and $G_3$), corresponding to time intervals [-3.8, 67.8]~s, [622.7, 664.7]~s, and [838.8, 1070.2]~s. 

The first episode with duration $t^{\rm pulse}_{90} \approx 88$ s starting at $T + T_{0,{\rm KW}} = -8.9$ s, the second episode with $t^{\rm pulse}_{90} \approx 106$ s beginning at $T + T_{0,{\rm KW}} = 609$ s (coinciding with the second GBM trigger), and the third episode with $t^{\rm pulse}_{90} \approx 477$ s starting at $T + T_{0,{\rm KW}} = 883$ s, where times are referenced to the KW trigger. The measured redshift is $z = 1.0924 \pm 0.0004$. The quiescent period between the first and second episodes spans approximately $\Delta t_{\rm gap} \sim 555$~s in the observer frame, corresponding to $\Delta t^{'}_{\rm gap} \sim 265$~s in the source rest frame \citep{Virgili2013}.

\begin{figure}[ht!]
{\bf a}\includegraphics[width=0.45\columnwidth]{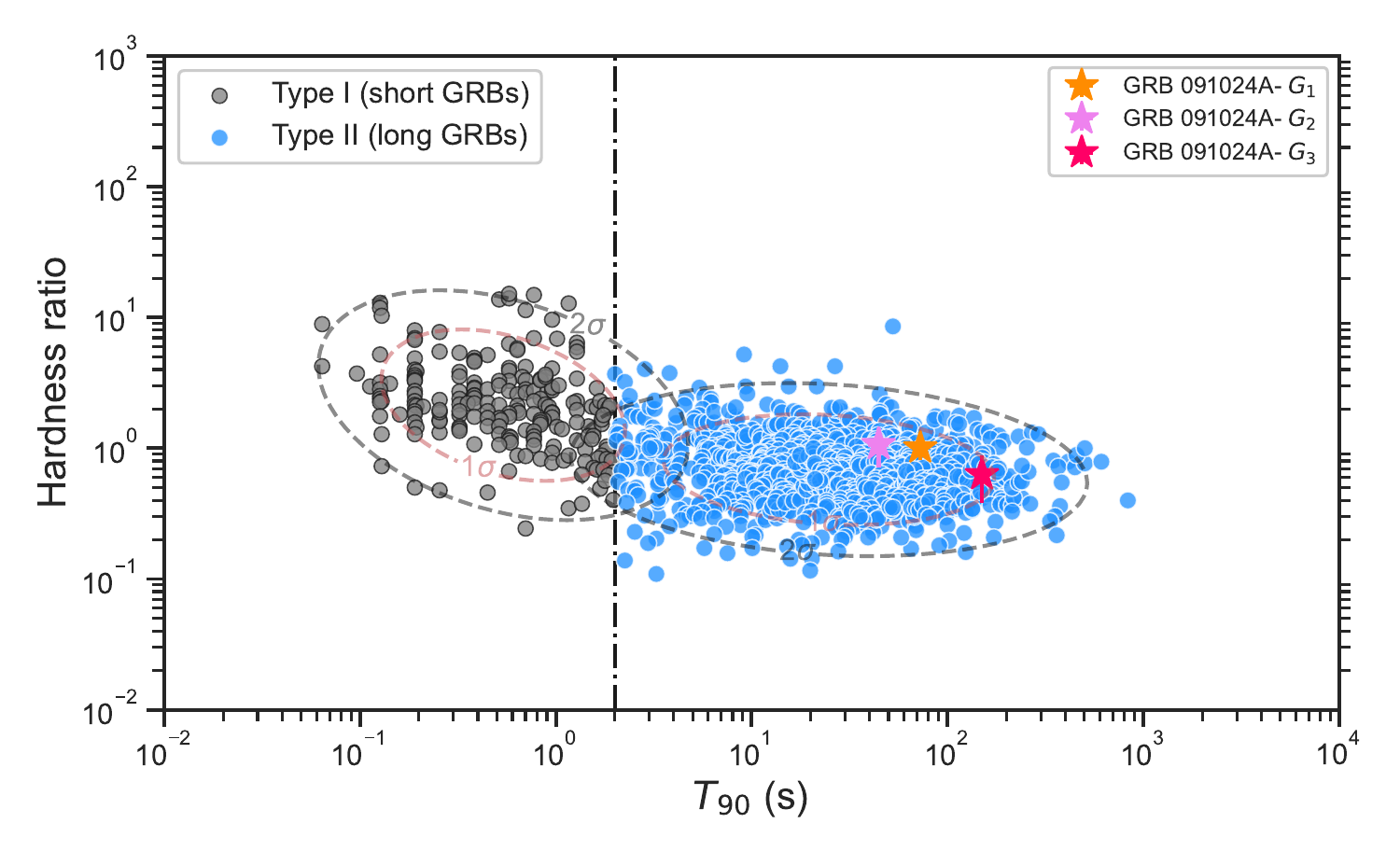}
{\bf b}\includegraphics[width=0.45\columnwidth]{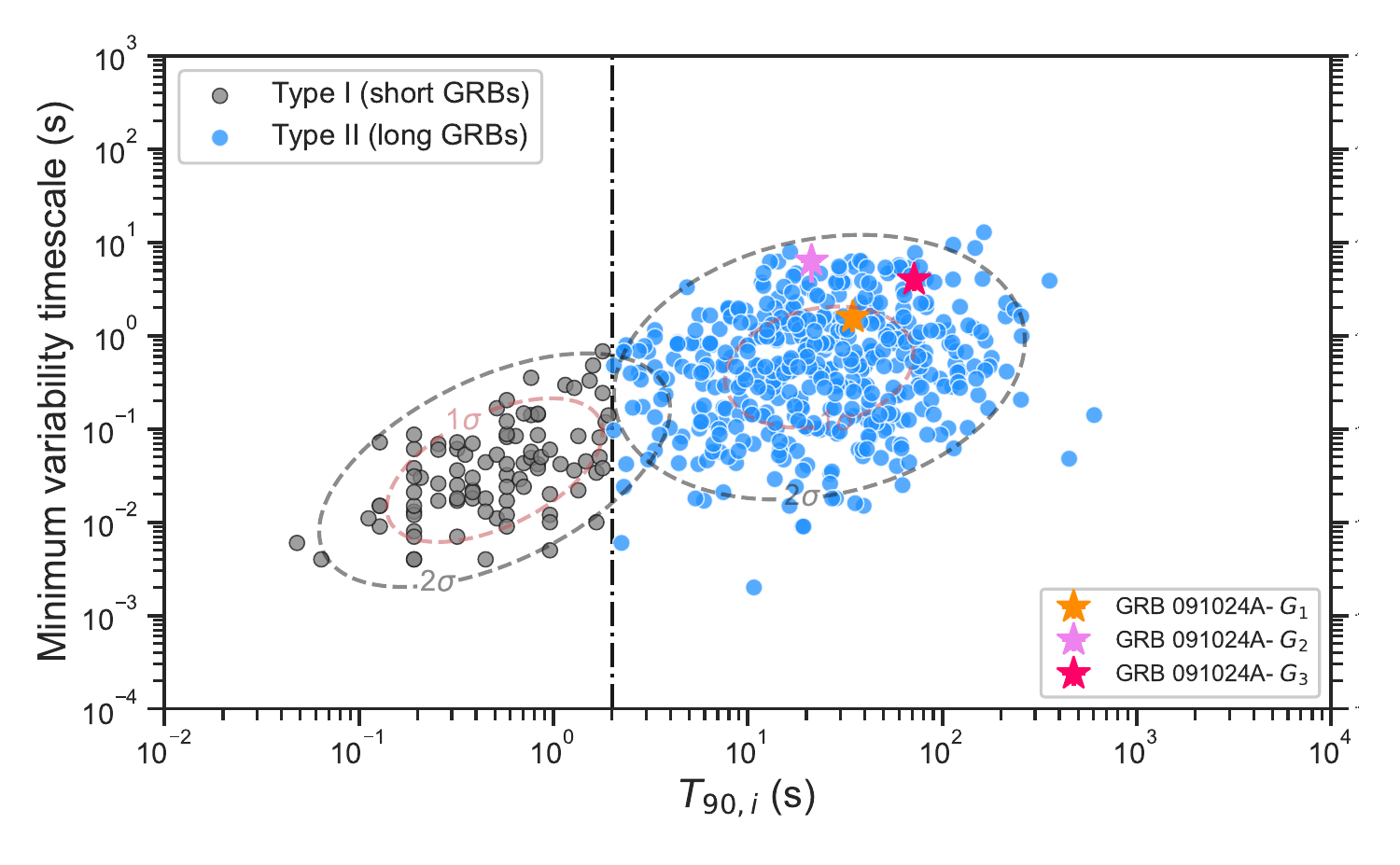}
{\bf c}\includegraphics[width=0.45\columnwidth]{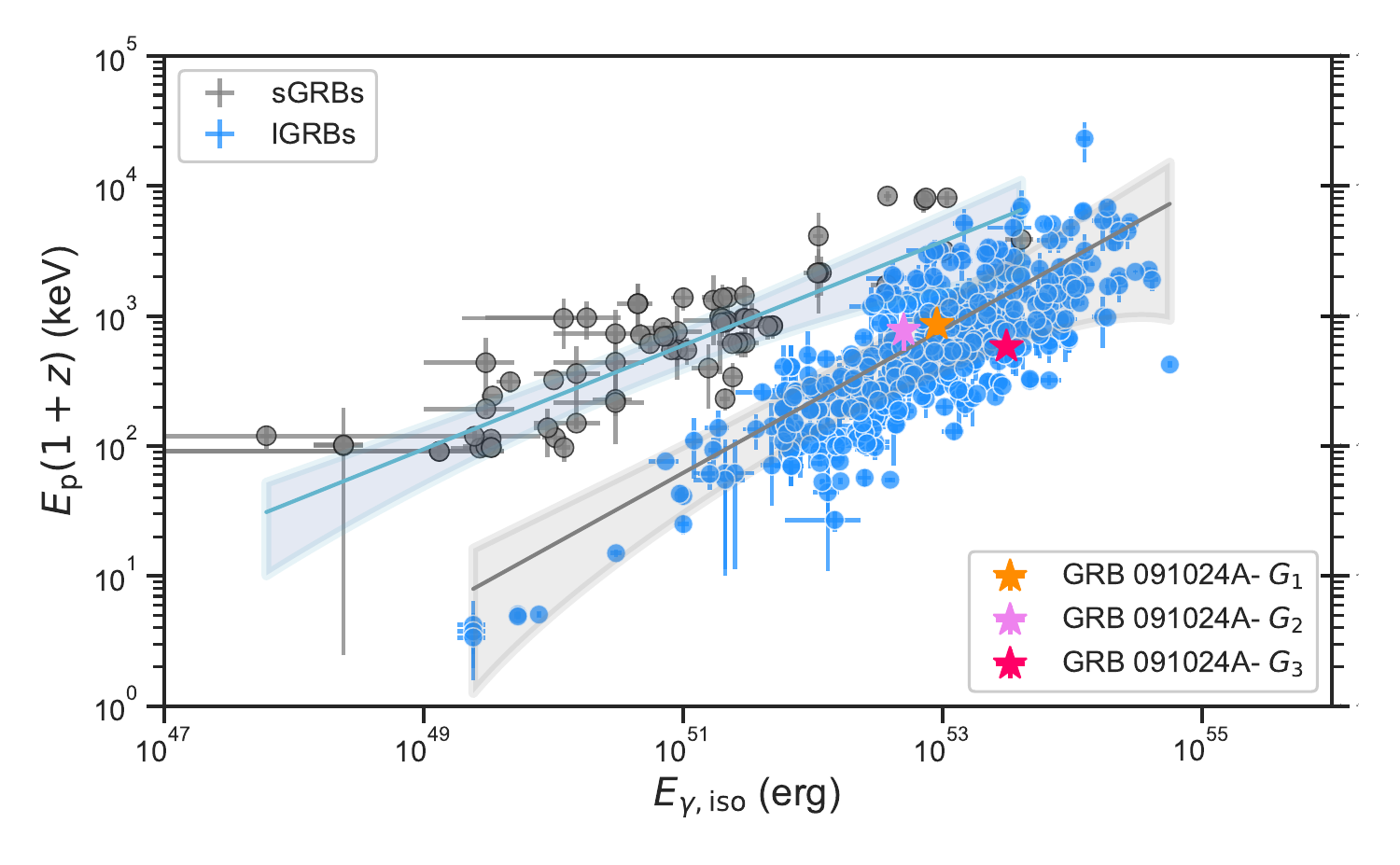}
{\bf d}\includegraphics[width=0.45\columnwidth]{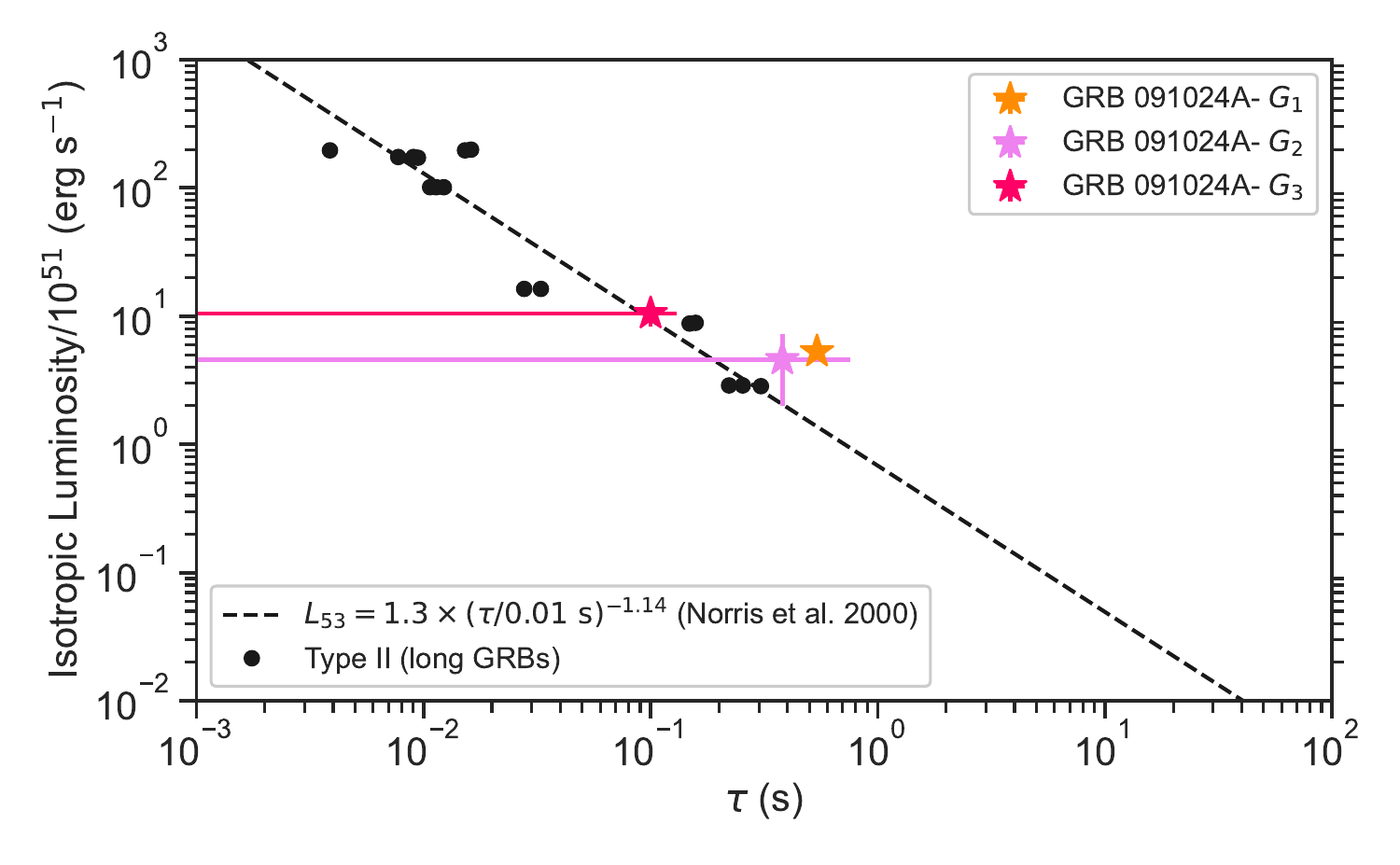}
{\bf e}\includegraphics[width=0.45\columnwidth]{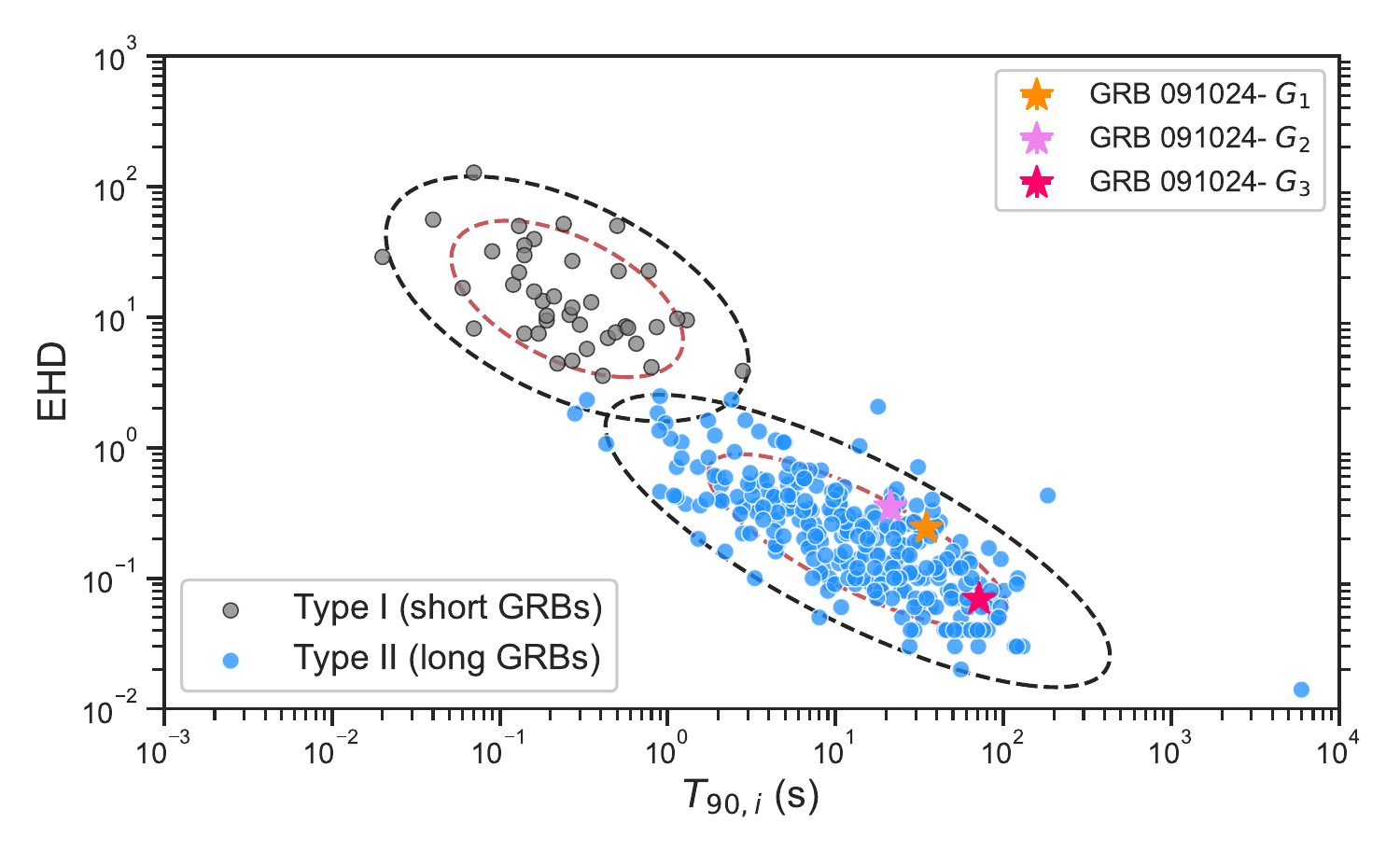}
{\bf f}\includegraphics[width=0.45\columnwidth]{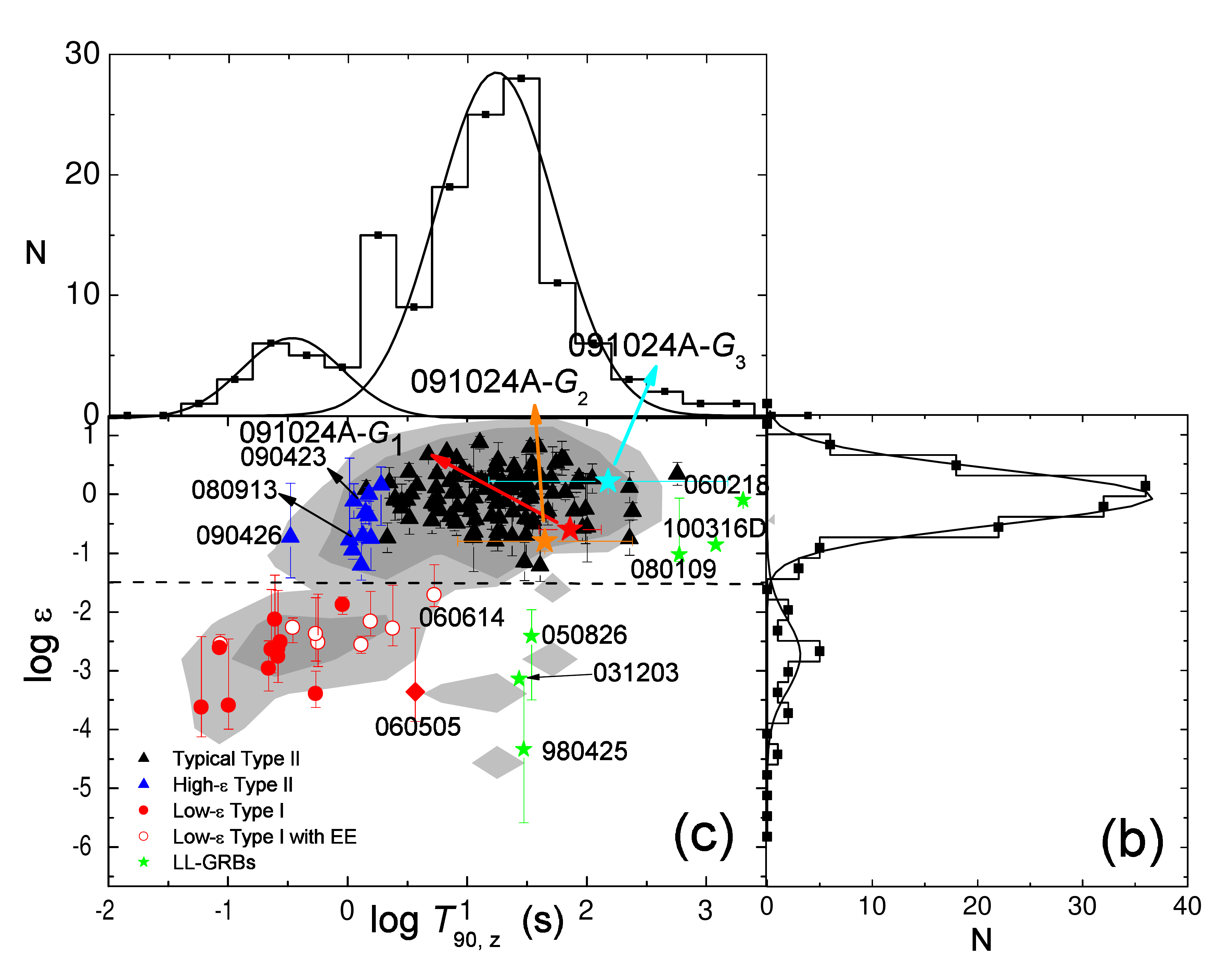}
\caption{{\bf GRB 091024A in traditional GRB classification scheme.} {\bf a-f}, The traditional GRB classification is illustrated based on the duration/hardness ratio diagram {\bf (a)}, the duration/MVT diagram {\bf (b)}, the Amati relation {\bf (c)}, the Norris relation {\bf (d)}, the duration/EHD diagram {\bf (e)}, and the duration/$\xi$ diagram {\bf (f)}. The \emph{Swift} (or \emph{Fermi}) GRB sample is represented by solid yellow and cyan points for the Type I (short) and Type II (long) burst populations, respectively. The three individual bursts of GRB 091024A are highlighted by red, orange and cyan stars. Elliptical dotted lines in different colors indicate the 1$\sigma$ and 2$\sigma$ regions of the bivariate normal distributions for the Type-I (short) and Type-II (long) burst populations. Bivariate distributions, using kernel density estimation, are shown in different shaded regions for the Type I (short) and Type II (long) burst populations. The traditional separation line ($t_{90}=2$ s) between short and long GRBs is indicated by a black dashed vertical line. Duration is plotted in the source frame for {\bf a, e, f} and the observed frame for {\bf b}.}
\label{fig:091024A_Classification}
\end{figure}

We measure timing and spectral descriptors and evaluate their consistency with long-GRB class in standard classification diagrams (Figures~\ref{fig:091024A_Classification}; Table~\ref{tab:091024A}). In Figure \ref{fig:091024A_Classification}a-f, we over-plot each individual sub-bursts observed in GRB 091024A on the hardness ratio-duration, MVT-duration, $E_{\rm p,z}$-$E_{\gamma,\rm iso}$, $L_{\rm p,iso}-\tau$ diagrams, EHD-duration, and $\varepsilon$-duration. Our analysis reveals that all sub-bursts consistently classify as Type~II across all diagnostic planes.

All three episodes have durations firmly in the long GRB regime: $T_{90} = 72.6 \pm 1.8$~s ($G_1$), $44.5 \pm 5.4$~s ($G_2$), and $150.0 \pm 10$~s ($G_3$). The hardness ratios $S(50\text{--}300~\rm keV)/S(10\text{--}50~\rm keV)$ are $1.01 \pm 0.20$, $1.07 \pm 0.36$, and $0.63 \pm 0.25$, respectively. MVT values are $1577 \pm 477$~ms, $6340 \pm 2661$~ms, and $3987 \pm 1009$~ms. Spectral lags (100--300~keV relative to 25--50~keV) are all positive: $(540 \pm 13)$~ms, $(380^{+380}_{-580})$~ms, and $(100^{+30}_{-130})$~ms. These temporal indicators place all episodes squarely within the Type~II (long GRB) distribution in both the $T_{90}$--hardness ratio and $T_{90}$--MVT planes.

Spectral analysis reveals peak energies of $E_{\rm p} = 412^{+69}_{-53}$~keV, $371^{+111}_{-71}$~keV, and $278^{+22}_{-18}$~keV for the three episodes, with photon indices $\alpha = -0.92 \pm 0.07$, $-1.17 \pm 0.07$, and $-1.38 \pm 0.02$ (compiled from \citealt{Gruber2011}). In the 1--$10^4$~keV band, the isotropic energies are $E_{\gamma,\rm iso} = (9.0^{+0.4}_{-0.3}) \times 10^{52}$~erg, $(5.0^{+0.4}_{-0.5}) \times 10^{52}$~erg, and $(3.1^{+0.07}_{-0.06}) \times 10^{53}$~erg, with peak luminosities of $(5.3 \pm 0.7) \times 10^{51}$~erg~s$^{-1}$, $(4.6 \pm 2.6) \times 10^{51}$~erg~s$^{-1}$, and $(1.05 \pm 0.22) \times 10^{52}$~erg~s$^{-1}$.

When projected onto the Amati and Norris lag--luminosity relations, all three episodes ($G_1$--$G_3$) lie within the $E_{\rm p,z}$--$E_{\gamma,\rm iso}$ distribution characteristic of long GRBs and are consistent with the Norris relation \citep{Norris2000}. Based on composite indicators (including $\varepsilon$ and $E_{\rm HD}$ parameters), all three episodes classify as Type~II (long GRB) under all diagnostic schemes.

\subsection{GRB 110709B}

\noindent{\bf Observations}.
GRB~110709B triggered the Burst Alert Telescope (BAT; \citealt{Barthelmy2005}) on board \emph{Swift} \citep{Gehrels2004} twice. The first triggered BAT (trigger=456967) at 21:32:39 UT on 2011 July 9 ($T_{0}$; \citealt{2011GCN.12122....1C}). Remarkably, BAT triggered a second time (trigger=456969) at 21:43:25 UT, about $11$ minutes ($t_{\rm gap} \simeq 650$ s) after the first trigger, and localized the second event to a consistent sky position \citep{2011GCN.12124....1B}. 

\emph{Swift} slewed promptly. The X-Ray Telescope (XRT; \citealt{Burrows2005a}) and the Ultraviolet/Optical Telescope (UVOT; \citealt{Roming2005}) began follow-up observations at $T_0+80.5$~s and $T_0+91$~s, respectively. XRT detected a bright X-ray afterglow and reported a refined position \citep{2011GCN.12130....1B} consistent with the BAT localization, while optical counterpart was identified within the XRT error region \citep{2011GCN.12140....1H,2011GCN.12157....1H}. 

\begin{figure}[ht!]
\centering
{\bf a}\includegraphics[width=1.0\textwidth]{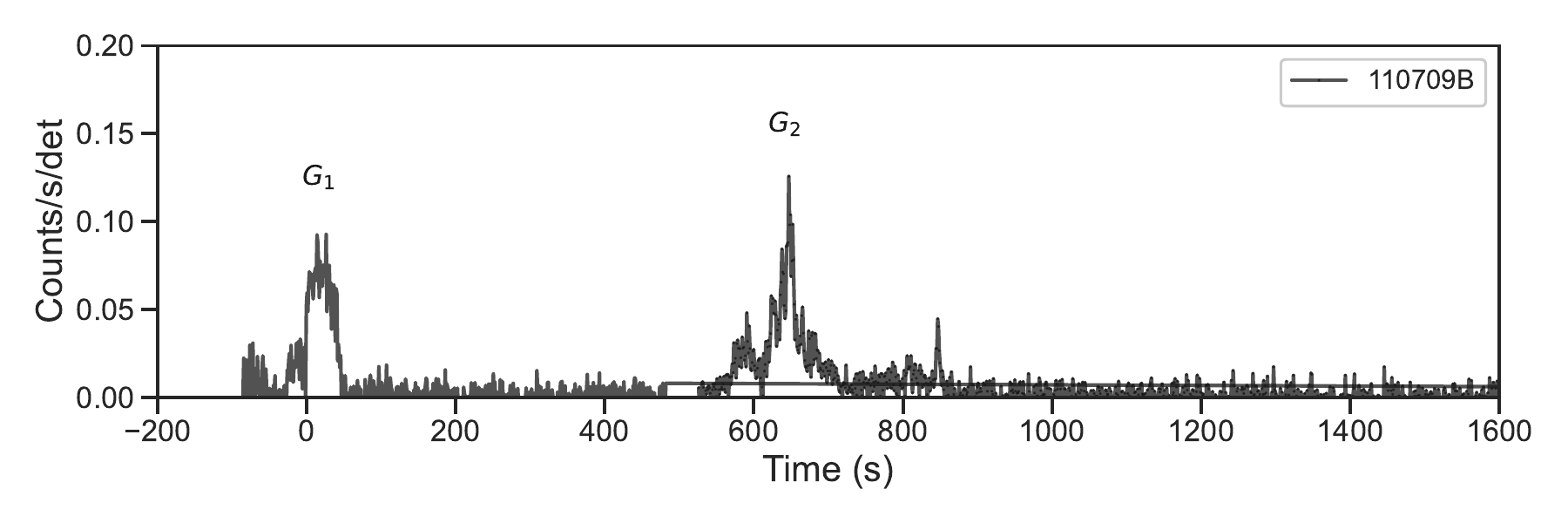}
{\bf b}\includegraphics[width=0.9\textwidth]{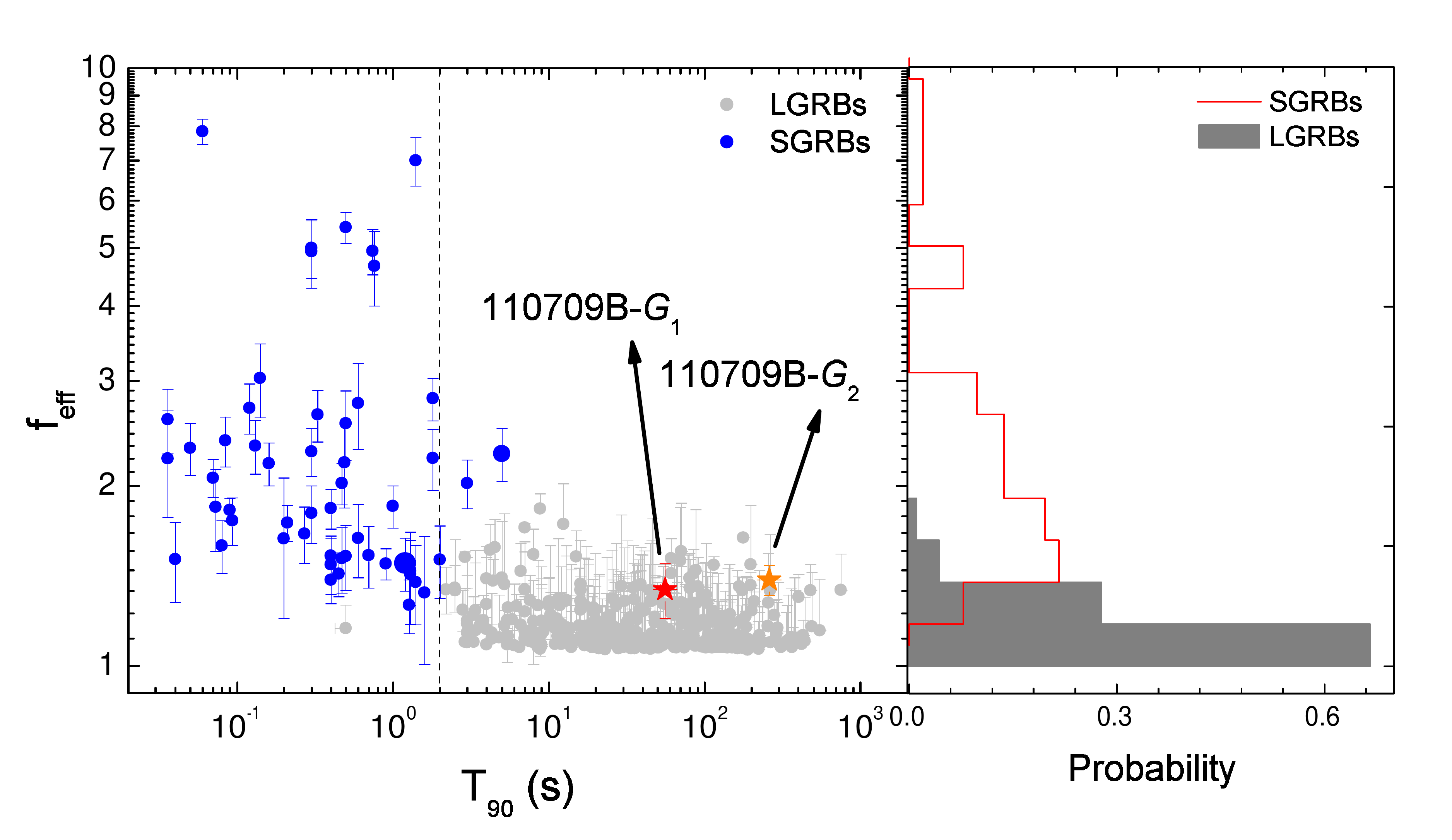}
\caption{Same as Figure \ref{fig:091024_lc} but for GRB 110709B. \label{fig:110709B_lc}}
\end{figure}

\noindent{\bf Pulse-resolved or Sub-burst Classification and Characteristics}.
We treat the two triggers as two emission episodes from a single progenitor and denote them as $G_1$ and $G_2$. We measure duration, hardness, minimum variability timescale, spectral lag, and spectral parameters for each sub-burst and compare their locations in standard classification planes with the \emph{Swift} and \emph{Fermi} GRB population (Figures~\ref{fig:110709B_lc}; Table~\ref{tab:110709B}).
We use these pulse-wise diagnostics, together with the $f$-parameter and other standard classification planes (Section~\ref{sec:Methodology}), to compare each sub-burst with the canonical long and short GRB populations.

Using the background-subtracted BAT 15--350~keV light curve (Figure~\ref{fig:110709B_lc}), we define the source intervals as $t_1$--$t_2=-28$ to 55~s ($G_1$) and $t_1$--$t_2=550$ to 865~s ($G_2$) relative to the first BAT trigger, corresponding to a quiescent gap of $\Delta t_{\rm gap}\simeq 494$~s in the observer frame \citep{ZhangBB2012}. The durations are $T_{90}=55.6\pm3.2$~s ($G_1$) and $T_{90}=259.2\pm8.8$~s ($G_2$), both placing firmly within the Type~II (long GRB) distribution in the $T_{90}$--hardness ratio and $T_{90}$--MVT planes. Both episodes show comparable prompt-emission properties in standard timing and spectral diagnostics, including hardness ratios $S_{50-100\,{\rm keV}}/S_{25-50\,{\rm keV}}=0.85\pm0.07$ ($G_1$) and $0.77\pm0.06$ ($G_2$), minimum variability timescales of $225\pm65$~ms ($G_1$) and $144\pm77$~ms ($G_2$), and positive spectral lags (50--100~keV relative to 25--50~keV) of $201\pm52$~ms ($G_1$) and $107\pm77$~ms ($G_2$). GRB~110709B lacks a measured redshift, preventing the use of rest-frame correlations such as the Amati relation. However, both episodes show consistent properties with long GRBs in all available observer-frame diagnostics.

We processed the BAT data using standard \texttt{HEAsoft} tools (version 6.10). Spectral fits using a cutoff power-law form yield characteristic energies of $E_c = 182 \pm 53$~keV and $116 \pm 9$~keV, with photon indices $\alpha = -1.09 \pm 0.09$ and $-1.12 \pm 0.01$. The fluences in the BAT band (15--150~keV) are $(8.95^{+0.29}_{-0.62}) \times 10^{-6}$~erg~cm$^{-2}$ and $(1.34^{+0.05}_{-0.07}) \times 10^{-6}$~erg~cm$^{-2}$.

\subsection{GRB 220627A}

\noindent{\bf Observations}. 
GRB~220627A triggered \emph{Fermi}/GBM twice (double-trigger event). GBM first triggered at 21:21:00.086 UT on 2022 June 27 (GRB~2220627890; hereafter $t_0$) and triggered a second time at 21:36:56.390 UT (GRB~2220627901), with a trigger separation of $\Delta t = 956.304$ s ($\sim$16 minutes) \citep{HuangYY2022}. The unusual temporal structure has led to two competing interpretations: either a gravitationally lensed burst or an ultra-long GRB with total duration of $\sim$1000 s \citep{2022GCN.32288....1R, HuangYY2022}.

\begin{figure}[ht!]
\centering
{\bf a}\includegraphics[width=1.0\textwidth]{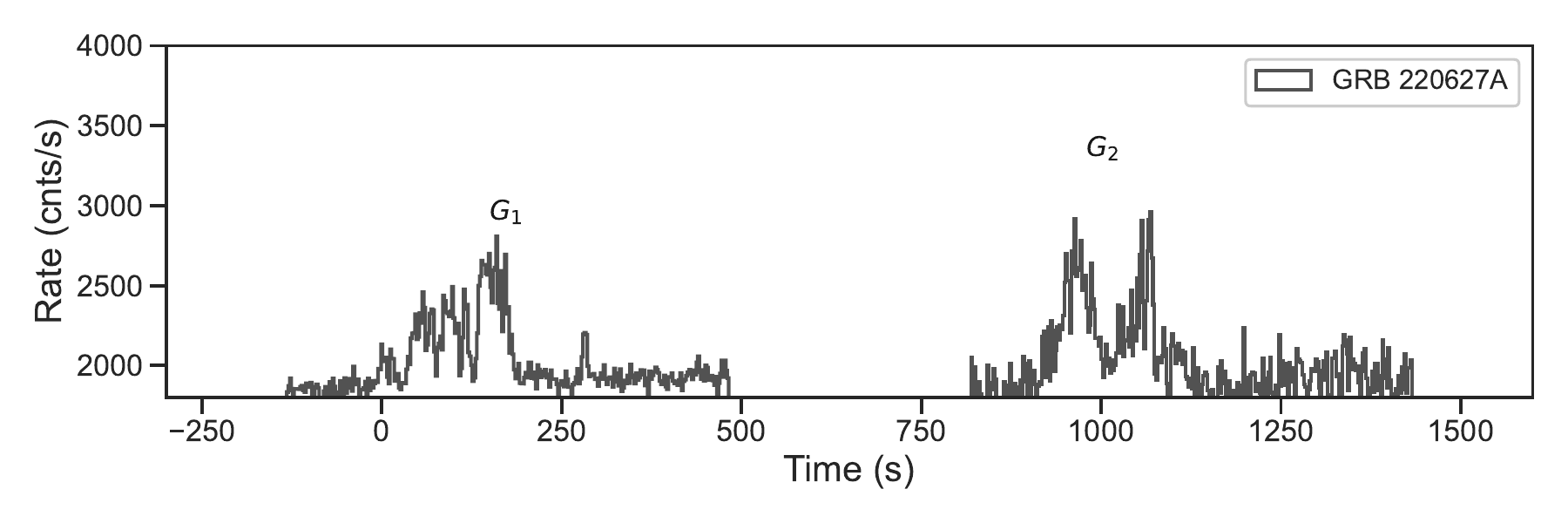}
{\bf b}\includegraphics[width=0.9\textwidth]{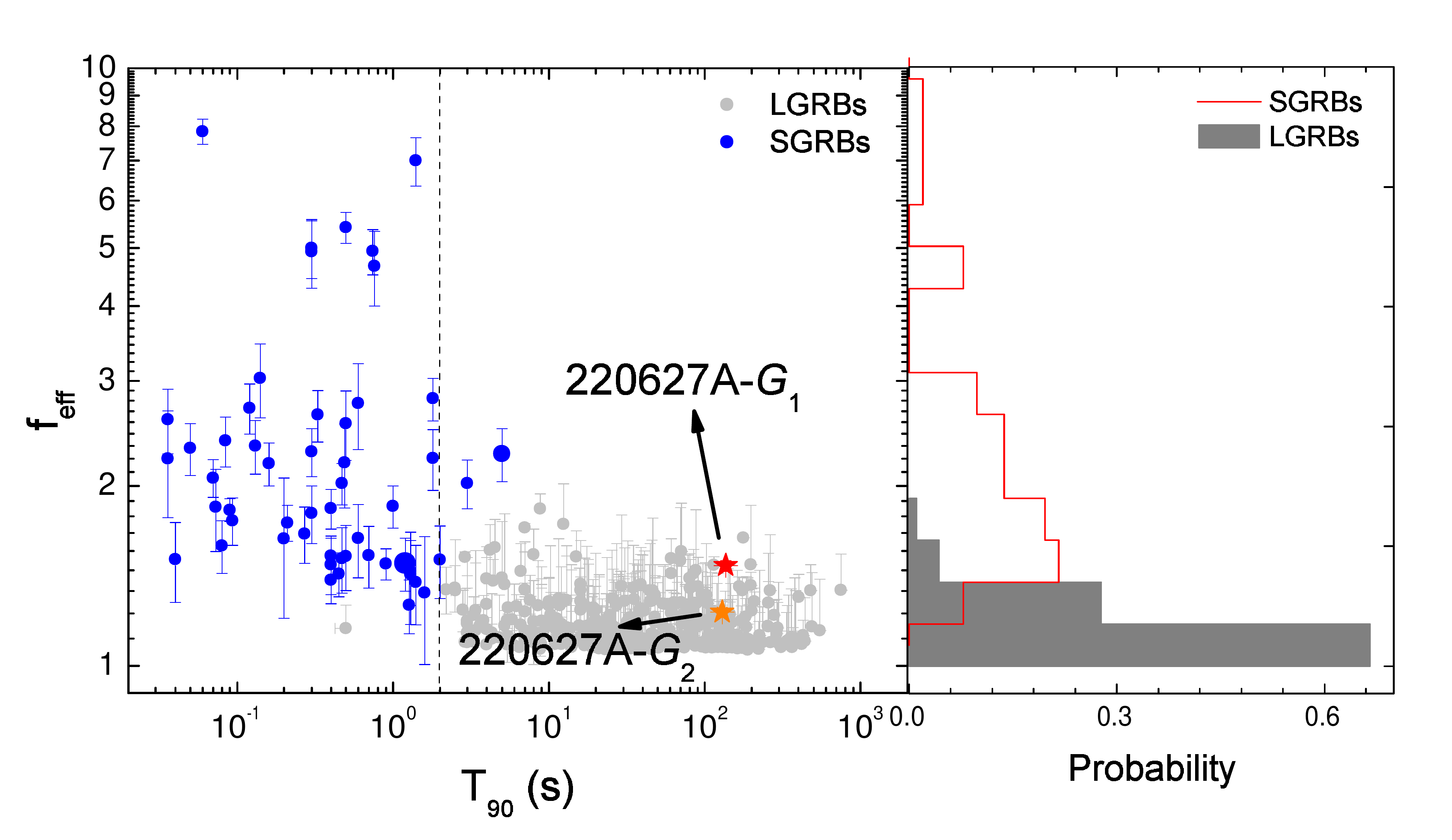}
\caption{Same as Figure \ref{fig:091024_lc} but for GRB 220627A.  \label{fig:220627A_lc}}
\end{figure}

The burst was detected by multiple instruments across a broad energy range, including the \textit{Fermi} Large Area Telescope
\citep[LAT;][]{diLalla2022}, \emph{Konus-Wind} \citep{Frederiks2022}, and \emph{Swift}/BAT-GUANO \citep{Raman2022}. LAT detected high-energy photons during the first episode and provided a localization enabling the identification of multi-wavelength afterglow emission \citep{diLalla2022,2022GCN.32288....1R}. Because the two triggers are consistent in sky position and both episodes show comparable prompt characteristics, GRB~220627A has been interpreted as either a gravitationally lensed event or an intrinsically ultra-long burst \citep{2022GCN.32288....1R,deWet2023}. We treat the two triggers as sub-bursts ($G_1$ and $G_2$) from the same event and analyze each episode separately, measuring their durations, hardness ratios, minimum variability timescales, spectral lags, and prompt spectral parameters (Figures~\ref{fig:220627A_lc}; Table~\ref{tab:220627A}).

\noindent{\bf Pulse-resolved or Sub-burst Classification and Characteristics}.
The prompt emission exhibits two clearly separated episodes ($G_1$ and $G_2$). The duration ($T_{90}$) is approximately 138 s in the 10--1000 keV range for $G_1$, while $G_2$ has $T_{90} \approx 127$ s over the same energy band \citep{deWet2023}. 
The striking similarity in temporal morphology and spectral properties of the two episodes initially suggested a gravitational lensing scenario, wherein the second episode represents a time-delayed, magnified image of the first \citep{2022GCN.32288....1R}. However, if the second episode represents continued central-engine activity rather than lensing, GRB~220627A would qualify as an ultra-long GRB given its extreme total duration exceeding 1000~s. The measured redshift is $z = 3.08$, with the quiescent period spanning approximately $\Delta t_{\rm q} \sim 600$~s in the observer frame and $\Delta t^{'}_{\rm q} \sim 147$~s in the rest frame \citep{deWet2023}.

Both episodes have durations characteristic of long GRBs: $T_{90} = 136.71 \pm 1.28$~s ($G_1$) and $126.98 \pm 8.84$~s ($G_2$), placing both episodes within the Type~II distribution in the $T_{90}$--hardness ratio and $T_{90}$--MVT planes. The hardness ratios $S(50\text{--}300~\rm keV)/S(10\text{--}50~\rm keV)$ are $1.74 \pm 0.33$ and $0.76 \pm 0.20$. MVT values are $3715 \pm 1841$~ms and $15691 \pm 14138$~ms, with spectral lags (100--300~keV relative to 25--50~keV) both positive: $(162.5 \pm 82.6)$~ms and $(184.1 \pm 91.9)$~ms. 

Spectral analysis yields peak energies of $E_p = 334.08 \pm 33.32$~keV and $327.94 \pm 58.84$~keV, with photon indices $\alpha = -0.89 \pm 0.04$ and $-0.88 \pm 0.07$. In the 1--$10^4$~keV band, the isotropic energies are $E_{\gamma,\rm iso} = (4.92^{+0.46}_{-0.42}) \times 10^{53}$~erg and $(1.78^{+0.13}_{-0.09}) \times 10^{54}$~erg, with peak luminosities (1~s) $(8.98^{+1.15}_{-1.25}) \times 10^{52}$~erg~s$^{-1}$ and $(3.68^{+0.74}_{-1.33}) \times 10^{52}$~erg~s$^{-1}$. When projected onto the Amati and Norris planes, both episodes align with the empirical distributions of long GRBs and classify as Type~II under the $\varepsilon$ and $E_{\rm HD}$ composite criteria.

\section{Discussion} \label{sec:discussion}

\subsection{Statistical Position and Timescales of Long Quiescent Periods}

The quiescent periods in all three double-trigger events span several hundred seconds in the observer frame: GRB~091024A, GRB~110709B, and GRB~220627A have $\Delta t_{\rm gap} \sim 555$~s, $\sim$494~s, and $\sim$600~s, respectively. For the two events with measured redshifts, the rest-frame quiescent periods are $\Delta t^{'}_{\rm gap} = \Delta t_{\rm gap}/(1+z)$, corresponding to $\Delta t^{'}_{\rm gap} \sim 265$~s for GRB~091024A and $\Delta t^{'}_{\rm gap} \sim 147$~s for GRB~220627A. Thus, double-trigger events exhibit not only extremely long quiescent periods in the observer frame but also maintain gaps of $\sim$100~s in the rest frame.

Within individual events, GRB~091024A's prompt emission divides into three distinct episodes ($G_1$, $G_2$, $G_3$), yet the $T_{90}$, hardness ratio, MVT, and spectral lag of each episode all fall within the Type~II (long GRB) distribution. Similarly, both episodes of GRB~220627A consistently fall into the Type~II region across the same classification planes. This indicates that long quiescent periods do not signal a change in burst type, but rather represent strong temporal segmentation within the same physical category.

\begin{figure}[ht!]
\centering
\includegraphics[width=1.0\textwidth]{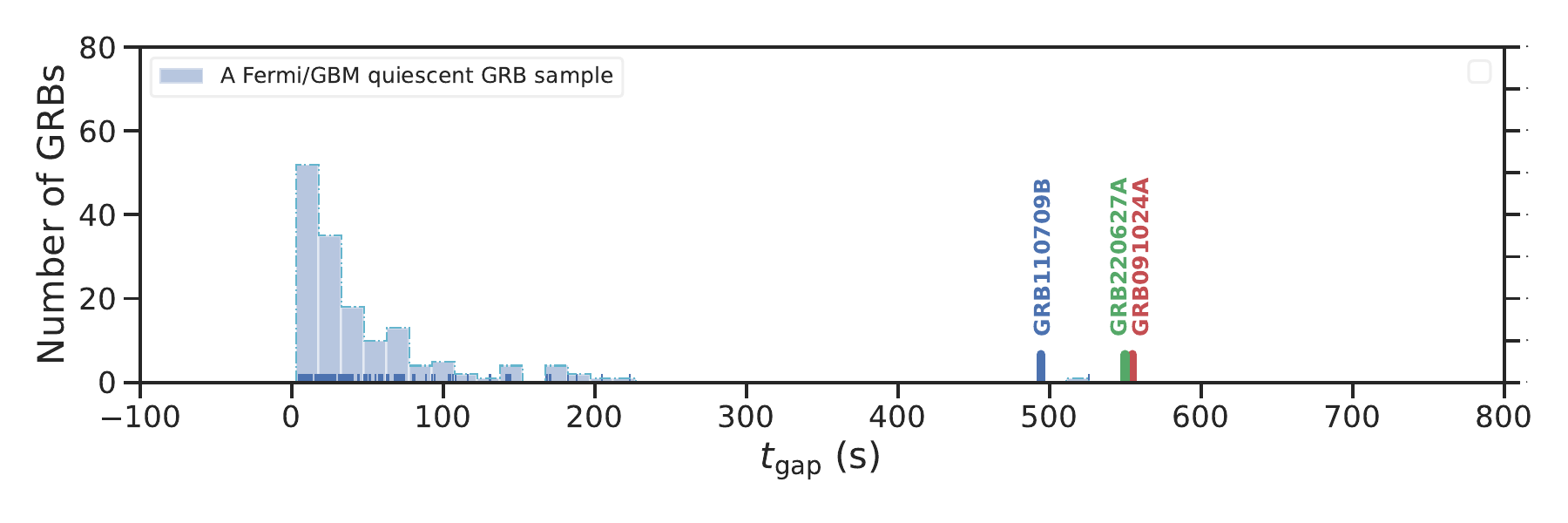}
\caption{{\bf Distribution of quiescent time intervals ($t_\mathrm{gap}$) for a combined sample of 154 GRBs (102 Fermi/GBM bursts and 52 Swift/BAT bursts) based on data collected from \cite{LanLin2018,LiLiande2022}.} The histogram shows that the vast majority of quiescent intervals are short, typically under a few hundred seconds, with most lying below $\sim$200 s. The frequency of longer gaps drops off steeply, indicating that extended dormant periods are uncommon. Only a few GRBs exhibit exceptionally long quiescent episodes: notably GRB 091024A, GRB 110709B, and GRB 220627A stand out with $\Delta t_\mathrm{gap}\sim 494$–600 s, placing them at the extreme long-duration tail of the distribution. These three events are clear outliers, highlighting that such prolonged pauses in prompt emission are exceedingly rare in the GRB population.\label{fig:t_gap}}
\end{figure}

To assess the uniqueness of the quiescent gap in GRB 091024A, GRB 110709B and GRB 220627A, we compiled a comprehensive dataset of quiescent GRBs by combining data from two independent studies \citep{LanLin2018,LiLiande2022}. These studies systematically analyzed quiescent times from ``quiescent" GRB samples detected by Swift and Fermi. Our compiled dataset includes 102 Fermi-detected GRBs \citep{LanLin2018} and 52 Swift-detected GRBs \citep{LiLiande2022}. Figure~\ref{fig:t_gap} presents the distribution of quiescent times ($\Delta t_{\rm gap}$) within this dataset. 
The quiescent periods of our three double-trigger events ($\sim$494--600~s in the observer frame) lie at the long-duration tail of this distribution, indicating they represent extreme cases of GRB quiescent periods.

\subsection{Connection and Differences with Ultra-Long GRBs}

Ultra-long GRBs typically exhibit sustained activity timescales of $10^3$--$10^4$~s, with observational signatures including continuous $\gamma$-ray emission, multi-segment prompt emission, or long intervals between prompt emission and subsequent high-energy activity. Double-trigger events similarly show activity spans on $\sim$1000~s timescales; for example, GRB~091024A's three emission episodes unfold over this duration, and GRB~220627A's two episodes span an even longer observation window. When characterized solely by activity span, double-trigger events phenomenologically overlap with ultra-long GRBs.

However, a key distinction lies in observational structure and trigger mechanism. The definition of double-trigger events depends directly on two independent triggers, requiring that the quiescent period be sufficiently long and interval emission fall below the trigger threshold, allowing the detector to return to regular monitoring mode after the first trigger and enabling a second trigger. In contrast, many ultra-long GRBs do not require a second trigger, either because detectable low-level activity persists during the quiescent period, or because the high-time-resolution data mode after triggering lasts long enough to continuously record subsequent emission as part of the same triggered event. Double-trigger events can thus be viewed as a subset of multi-episode long GRBs at the observational trigger level, statistically biased toward cases with exceptionally long quiescent periods and extremely weak interval emission.

\subsection{Origin of Double-Triggering: Observational Selection Effects and Physical Processes}

\subsubsection{Influence of Trigger Conditions on Double-Trigger Selection Effects}

The formation of double-trigger events involves explicit instrumental selection conditions. Generally, trigger systems make decisions based on count-rate significance across multiple time windows and energy bands. After triggering, systems enter high-time-resolution data mode and return to regular monitoring under certain conditions. If subsequent emission arrives while the instrument remains in post-trigger data mode, or if persistent low-level emission during the interval prevents count rates from returning to background statistical levels, subsequent emission typically will not register as a second event but will be recorded as post-trigger emission from the same trigger. Conversely, only when the quiescent period is sufficiently long and interval emission sufficiently weak will both conditions, trigger reset and re-exceeding trigger threshold, be simultaneously satisfied, forming a double-trigger observational result.

This selection effect explains two phenomenological facts. First, GRB~091024A's prompt emission divides into three segments, yet only two triggers occurred, suggesting that the trigger mechanism does not necessarily produce a new trigger for every emission re-brightening, but is jointly determined by trigger windows and background recovery conditions. Second, the three double-trigger events lie at the long-duration tail of the $t_{\rm gap}$ distribution, consistent with the expectation that only the most extreme quiescent periods are likely to produce second triggers.

Cases like GRB~220627A demonstrate that double-triggering depends largely on the satellite's trigger algorithm and background fitting time window. For ordinary long GRBs with extended quiescent periods, placement at different redshifts, observation under different signal-to-noise conditions, or changes to the satellite's background reset time threshold could result in their being recorded as double-trigger events.

Double-trigger GRBs thus do not constitute an independent physical subclass, but rather represent the extreme manifestation of temporal activity distribution in long GRBs.

\subsubsection{Physical Constraints: Are Sub-bursts Repeated Activity of the Same Central Engine?}

Although the definition of double-trigger includes observational selection effects, our results provide constraints on their physical nature. All sub-bursts from the three events consistently fall into the Type~II region under traditional classification diagnostics, including $T_{90}$--hardness ratio, $T_{90}$--MVT, and empirical relation tests involving spectra and temporal structure. For GRB~091024A, $G_1$--$G_3$ are all compatible with the Amati and Norris relations, and all spectral lags are positive and within the typical range for long GRBs. The two emission episodes of GRB~220627A show consistent behavior across the same planes. The interpretation most consistent with observational facts is that the two triggers correspond to segmented activity of the same long GRB central engine, rather than coincidental superposition of two different source events.

At the physical mechanism level, long quiescent periods require significant power reduction or energy release cessation from the central engine over $\sim$100~s timescales, followed by subsequent re-enhancement. Frameworks that can satisfy this requirement include, segmented accretion-powered processes (e.g., rapid decrease followed by increase in accretion rate) and staged changes in jet production or energy dissipation conditions. Since this work focuses primarily on prompt-phase temporal and spectral classification diagnostics, specific microphysical processes regarding accretion and jets require discussion in conjunction with subsequent multiwavelength observations and more detailed time-resolved spectral analysis. The direct conclusion from our current results is that the classification properties of different sub-bursts are highly consistent, providing empirical support for repeated restart of the same central engine.

Our analysis shows that each triggered event in double-trigger GRBs shows no significant differences from ordinary long GRBs in either temporal characteristics (such as pulse width and variability) or spectral characteristics (such as $E_{\rm p}$ evolution and the Amati relation). This strongly suggests they do not originate from special physical progenitor systems.

The double-triggering phenomenon more likely associates with the central engine (possibly a black hole accretion disk or millisecond magnetar) being in an unstable accretion or spin-down state. According to fallback accretion models, stellar envelope material does not all fall into the black hole at once after collapse, but forms intermittent accretion flows. When the accretion rate falls below a critical value temporarily, the jet is interrupted, forming a quiescent period. As fallback material arrives, the accretion rate rises again, the jet restarts, and a second trigger is produced.

\subsection{A Unified View of Long-Quiescent-Period Long GRBs}

Combining classification results and quiescent period distributions, this work supports a unified phenomenological picture: double-trigger events constitute a subclass of long GRBs with exceptionally long quiescent periods, with each triggered sub-burst consistent with typical Type~II long GRBs in duration, hardness, variability timescale, and spectral lag. The occurrence of double-triggering requires simultaneous satisfaction of physical conditions (existence of long quiescent periods and re-enhanced emission episodes) and observational conditions (trigger system reset after the quiescent period and subsequent emission re-exceeding trigger threshold). Double-triggering is thus more appropriately viewed as an observational marker of long-quiescent-period long GRBs rather than a criterion for new burst type classification.

\section{Conclusions} \label{sec:conclusions}

In this paper, we analyzed three GRB events independently triggered twice by \textit{Swift}/BAT or \textit{Fermi}/GBM (GRB~091024A, GRB~110709B, GRB~220627A). We treated each triggered emission episode as an independent sub-burst and, under a unified data processing and diagnostic framework, measured and classified each by duration, hardness, variability characteristics, spectral lag, spectral parameters, and energetics.

Our main conclusions are as follows:
\begin{enumerate}
\item All sub-bursts from the three events consistently fall within the Type~II (long GRB) distribution across multiple traditional classification planes. In duration, GRB~091024A's three emission episodes ($G_1$/$G_2$/$G_3$) have $T_{90} = 72.6 \pm 1.8$~s, $44.5 \pm 5.4$~s, and $150.0 \pm 10$~s; GRB~110709B's two episodes ($G_1$/$G_2$) have $55.6 \pm 3.2$~s and $259.2 \pm 8.8$~s; GRB~220627A's two episodes have $136.71 \pm 1.28$~s and $126.98 \pm 8.84$~s. The corresponding hardness ratios, MVT values, and spectral lags are all consistent with the long GRB population, with all spectral lags positive.

\item For the two events with measured redshifts (GRB~091024A: $z = 1.0924$; GRB~220627A: $z = 3.08$), each sub-burst is compatible with the empirical distributions of long GRBs in the $E_{\rm p,z}$--$E_{\gamma,\rm iso}$ (Amati) and $L_{\gamma,\rm p,iso}$--$\tau$ (Norris) planes. This indicates that even when multiple prompt emission episodes separated by long quiescent periods exist within the same event, their energetic scaling relations show no systematic deviation from the long GRB population.

\item The quiescent period timescales corresponding to the double-trigger phenomenon lie at the long-duration tail of the statistical distribution. The quiescent periods of the three events are approximately $\Delta t_{\rm gap} \sim 555$~s, $\sim$494~s, and $\sim$600~s in the observer frame. For the two events with redshifts, the corresponding rest-frame quiescent periods are approximately $\Delta t^{'}_{\rm q} \sim 265$~s (GRB~091024A) and $\sim$147~s (GRB~220627A). In a \textit{Fermi}/GBM quiescent GRB sample, these three events lie at the long-duration tail of the $t_{\rm gap}$ distribution, indicating that double-triggering serves more as an observational marker of extreme long-quiescent-period long GRBs rather than a distinct burst type.

\item Based on these results, we favor the following interpretation: double-trigger events in our sample are consistent with ordinary Type~II long GRBs in physical classification, with the two triggers more likely corresponding to segmented activity of the same central engine. Meanwhile, the formation of double-trigger events is inevitably influenced by trigger algorithms and observation mode switching, requiring simultaneous satisfaction of interval emission weak enough for trigger system reset and subsequent emission strong enough to re-exceed trigger threshold.
\end{enumerate}

Future work should proceed in two directions: first, systematically searching \textit{Swift}/BAT and \textit{Fermi}/GBM archives for multiple triggers with consistent positions and ultra-long quiescent periods without double-triggering events to expand the sample; second, performing more sensitive background modeling and weak signal detection during quiescent periods to determine whether persistent low-level emission exists during intervals, thereby assessing the degree to which trigger selection effects bias affects the double-trigger sample.

\acknowledgments
This work is supported by the Natural Science Foundation of China (grant No. 11874033), the KC Wong Magna Foundation at Ningbo University, and made use of the High Energy Astrophysics Science Archive Research Center (HEASARC) Online Service at the NASA/Goddard Space Flight Center (GSFC). The computations were supported by the high performance computing center at Ningbo University.

\vspace{15mm}
\facilities{\textit{Swift}, \textit{Fermi}}
\software{
{\tt 3ML} \citep{Vianello2015}, 
{\tt matplotlib} \citep{Hunter2007}, 
{\tt NumPy} \citep{Harris2020,Walt2011}, 
{\tt SciPy} \citep{Virtanen2020}, 
{\tt $lmfit$} \citep{Newville2016}, 
{\tt astropy} \citep{AstropyCollaboration2013},
{\tt pandas} \citep{Reback2022},
{\tt emcee} \citep{Foreman-Mackey2013},
{\tt seaborn} \citep{Waskom2017}}  
\vspace{75mm}
\bibliography{MyReferences.bib}

\clearpage
\begin{table*}[htbp]
\centering
\footnotesize
\setlength{\tabcolsep}{2pt} 
\caption{Individual burst properties of GRB 091024A \label{tab:091024A}}
\centering
\begin{tabular}{l|c|c|c|c|c|c}
\hline
& \multicolumn{2}{c|}{GRB 091024A-$G_1$} & \multicolumn{2}{c}{GRB 091024A-$G_2$} & \multicolumn{2}{c}{GRB 091024A-$G_3$}\\
\hline
&Value&Classification&Value&Classification\\
\hline
Time interval (10-1000 keV) [$t_1$$\sim$$t_2$]&[-3.8$\sim$67.8]~(s)&&[622.7$\sim$664.7]~(s)&&[838.8$\sim$1070.2]~(s)&\\
Duration (10-1000 keV) [$T_{90}$]&72.6$\pm$1.8~(s)&Long&44.5$\pm$5.4~(s)&Long&150.0$\pm$10~(s)&Long\\
Hardness ratio [$\frac{S(50-300 {\rm keV})}{S(10-50{\rm keV})}$]&1.01$\pm$0.20&Long&1.07$\pm$0.36&Long&0.63$\pm$0.25&Long\\
Minimum variability timescale (MVT)&1577$\pm$477~(ms)&Long&6340$\pm$2661~(ms)&Long&3987$\pm$1009~(ms)&Long\\
$f$- parameters [$F_{\rm p}/F_{\rm b}$]&1.18$\pm$0.04&Long&1.27$\pm$0.04&Long&1.33$\pm$0.04&Long\\
Time lag [100-300 keV $\sim$ 25-50 keV]&(540$\pm$13)~(ms)&Long&(380$^{+380}_{-580}$)~(ms)&Long&(100$^{+30}_{-130}$)~(ms)&Long\\
Spectral photon index [$\alpha$]&-0.92$\pm$0.07&&-1.17$\pm$0.07&&-1.38$\pm$0.02&\\
Spectral peak energy [$E_{\rm p}$]&412$^{+69}_{-53}$~(keV)&&371$^{+111}_{-71}$~(keV)&&278$^{+22}_{-18}$~(keV)&\\
Bolometric (1-10$^{4}$ keV) fluence [$S_{\gamma}$]~(erg~cm$^{-2}$)&(1.81$\pm$0.07)$\times$10$^{-5}$&&(0.79$\pm$0.04)$\times$10$^{-5}$&&(6.73$\pm$0.09)$\times$10$^{-5}$&\\
Peak luminosity (1-10$^{4}$ keV) [$L_{\rm \gamma, p}$]~(erg~s$^{-1}$)&(5.3$\pm$0.7)$\times$10$^{51}$&&(4.6$\pm$2.6)$\times$10$^{51}$&&(1.05$\pm$0.22)$\times$10$^{52}$&\\
Isotropic energy (1-10$^{4}$ keV) [$E_{\rm \gamma, iso}$]&(9.0$^{+0.4}_{-0.3}$)$\times$10$^{52}$(erg)&Long&(5.0$^{+0.4}_{-0.5}$)$\times$10$^{52}$)~(erg)&Long&(3.1$^{+0.07}_{-0.06}$)$\times$10$^{53}$&Long\\
$\epsilon$-parameter [$E_{\gamma,\rm iso,52}/E^{5/3}_{\rm p,z,2}$]&0.25$^{+0.07}_{-0.05}$&Long&0.16$^{+0.08}_{-0.05}$&Long&1.65$^{+0.22}_{-0.18}$&Long\\
EHD-parameter [$E_{\rm p,i,2} E^{-0.4}_{\rm iso,51}T^{-0.5}_{90,i}$]&0.24$^{+0.04}_{-0.03}$&Long&0.35$^{+0.11}_{-0.07}$&Long&0.07$^{+0.006}_{-0.005}$&Long\\
Amati relation [$E_{\rm p,z} \propto E^{0.5}_{\gamma,\rm iso}$]&On the relation&Long&On the relation&Long&On the relation&Long\\
Norris relation [$L_{\gamma,\rm p, iso}\propto \tau^{-1.14}$]&On the relation&Long&On the relation&Long&On the relation&Long\\
\hline
\end{tabular}
\end{table*}

\clearpage
\begin{table*}[htbp]
\centering
\footnotesize
\caption{Individual burst properties of GRB 110709B \label{tab:110709B}}
\centering
\begin{tabular}{l|c|c|c|c}
\hline
& \multicolumn{2}{c|}{GRB 110709B-$G_1$} & \multicolumn{2}{c}{GRB 110709B-$G_2$} \\
\hline
&Value&Classification&Value&Classification\\
\hline
Time interval (15-350 keV) [$t_1$$\sim$$t_2$]&[-28$\sim$55]~(s)&&[550.0$\sim$865.0]~(s)&\\
Duration (15-350 keV) [$T_{90}$]&55.6$\pm$3.2~(ms)&Long&259.2$\pm$8.8~(ms)&Long\\
Hardness ratio [$\frac{S(50-100 {\rm keV})}{S(25-50{\rm keV})}$]&0.85$\pm$0.07&Long&0.77$\pm$0.06&Long\\
Minimum variability timescale (MVT)&225$\pm$65~(ms)&Long&144$\pm$77~(ms)&Long\\
$f$- parameters [$F_{\rm p}/F_{\rm b}$]&1.34$\pm$0.14&Long&1.39$\pm$0.08&Long\\
Time lag [50-100 keV $\sim$ 25-50 keV]&(201$\pm$52)~(ms)&Long&(107$\pm$77)~(ms)&Long\\
Spectral photon index [$\alpha$]&-1.09$\pm$0.09&&-1.12$\pm$0.01&\\
Spectral peak energy [$E_{\rm c}$]&182$\pm$53~(keV)&&116$\pm$9~(keV)&\\
BAT (15-150 keV) fluence [$S_{\gamma}$]&(8.95$^{+0.29}_{-0.62}$)$\times$10$^{-6}$~(erg~cm$^{-2}$)&&(1.34$^{+0.05}_{-0.07}$)$\times$10$^{-6}$~(erg~cm$^{-2}$)&Long\\
\hline
\end{tabular}
\end{table*}

\clearpage
\begin{table*}[htbp]
\centering
\footnotesize
\caption{Individual burst properties of GRB 220627A \label{tab:220627A}}
\centering
\begin{tabular}{l|c|c|c|c}
\hline
& \multicolumn{2}{c|}{GRB 220627A-$G_1$} & \multicolumn{2}{c}{GRB 220627A-$G_2$} \\
\hline
&Value&Classification&Value&Classification\\
\hline
Time interval (10-1000 keV) [$t_1$$\sim$$t_2$]&[-1.28$\sim$138.99]~(s)&&[-8.84$\sim$135.82]~(s)&\\
Duration (10-1000 keV) [$T_{90}$] &136.71$\pm$1.28~(s)&Long&126.98$\pm$8.84~(s)&Long\\
Hardness ratio [$\frac{S(50-300 {\rm keV})}{S(10-50{\rm keV})}$]&1.74$\pm$0.33?&Long&0.76$\pm$0.20&Long\\
Minimum variability timescale (MVT)&3715$\pm$1841~(ms)&Long&15691$\pm$14138~(ms)&Long\\
$f$- parameters [$F_{\rm p}/F_{\rm b}$]&1.47$\pm$0.03&Long&1.23$\pm$0.02&Long\\
Time lag[100-300 keV $\sim$ 25-50 keV]&(162.5$\pm$82.6)~(ms)&Long&(184.1$\pm$91.9)~(ms)&Long\\
Spectral photon index [$\alpha$]&-0.89$\pm$0.04&&-0.88$\pm$0.07&\\
Spectral peak energy [$E_{\rm p}$]&334.08$\pm$33.32~(keV)&&327.936$\pm$58.84~(keV)&\\
Fluence [$S_{\gamma}$](10-1000 keV)&(5.54$\pm$0.01)$\times$10$^{-5}$~(erg~cm$^{-2}$)&&(1.08$\pm$0.02)$\times$10$^{-5}$~(erg~cm$^{-2}$)&\\
Peak luminosity (15-150 keV) [$L_{\rm \gamma, p, iso}$]&(8.98$^{+1.15}_{-1.25}$)$\times$10$^{52}$~(erg~s$^{-1}$)[1-s]&&(3.68$^{+0.74}_{-1.33}$)$\times$10$^{52}$~(erg~s$^{-1}$)[1-s]&\\
Isotropic energy (1-10$^{4}$ keV) [$E_{\rm \gamma, iso}$]&(4.92$^{+0.46}_{-0.42}$)$\times$10$^{53}$(erg)&Long&(1.78$^{+0.13}_{-0.09}$)$\times$10$^{54}$~(erg)&Long\\
$\epsilon$-parameter [$E_{\gamma,\rm iso,52}/E^{5/3}_{\rm p,z,2}$]&2.68$^{+0.45}_{-0.42}$&Long&1.80$^{+0.49}_{-0.41}$&Long\\
EHD-parameter [$E_{\rm p,i,2} E^{-0.4}_{\rm iso,51}T^{-0.5}_{90,i}$]&(1.01$^{+0.16}_{-0.14}$)$\times$10$^{-1}$&Long&(4.58$^{+0.23}_{-0.23}$)$\times$10$^{-1}$&Long\\
Amati relation [$E_{\rm p,z} \propto E^{0.5}_{\gamma,\rm iso}$]&On the relation&Long&On the relation&Long\\
Norris relation [$L_{\gamma,\rm p, iso}\propto \tau^{-1.14}$]&On the relation&Long&On the relation&Long\\
\hline
\end{tabular}
\end{table*}

\clearpage
\begin{table*}[htbp]
\centering
\footnotesize
\caption{Proprieties of Double-trigger Gamma-ray Bursts  \label{tab:proprieties}}
\centering
\begin{tabular}{l|ccccc}
\hline
GRB&$z$&Quiescent gap&Quiescent gap&Reference\\
\hline
&&($\Delta t_{\rm q}$)&($\Delta t^{'}_{\rm q}$)&\\
\hline
&&Observer-frame (s)&Rest-frame (s)\\
\hline
GRB 091024A&1.0924$\pm$0.0004&$\sim$555&$\sim$265&\cite{Virgili2013}\\
GRB 110709B&\nodata&$\sim$494&\nodata&\cite{ZhangBB2012}\\
GRB 220627A&3.08&$\sim$600&$\sim$147&\cite{deWet2023}\\
\hline
\end{tabular}
\end{table*}

\clearpage

\clearpage

\clearpage

\clearpage
\begin{figure}[ht!]
{\bf a}\includegraphics[width=0.45\columnwidth]{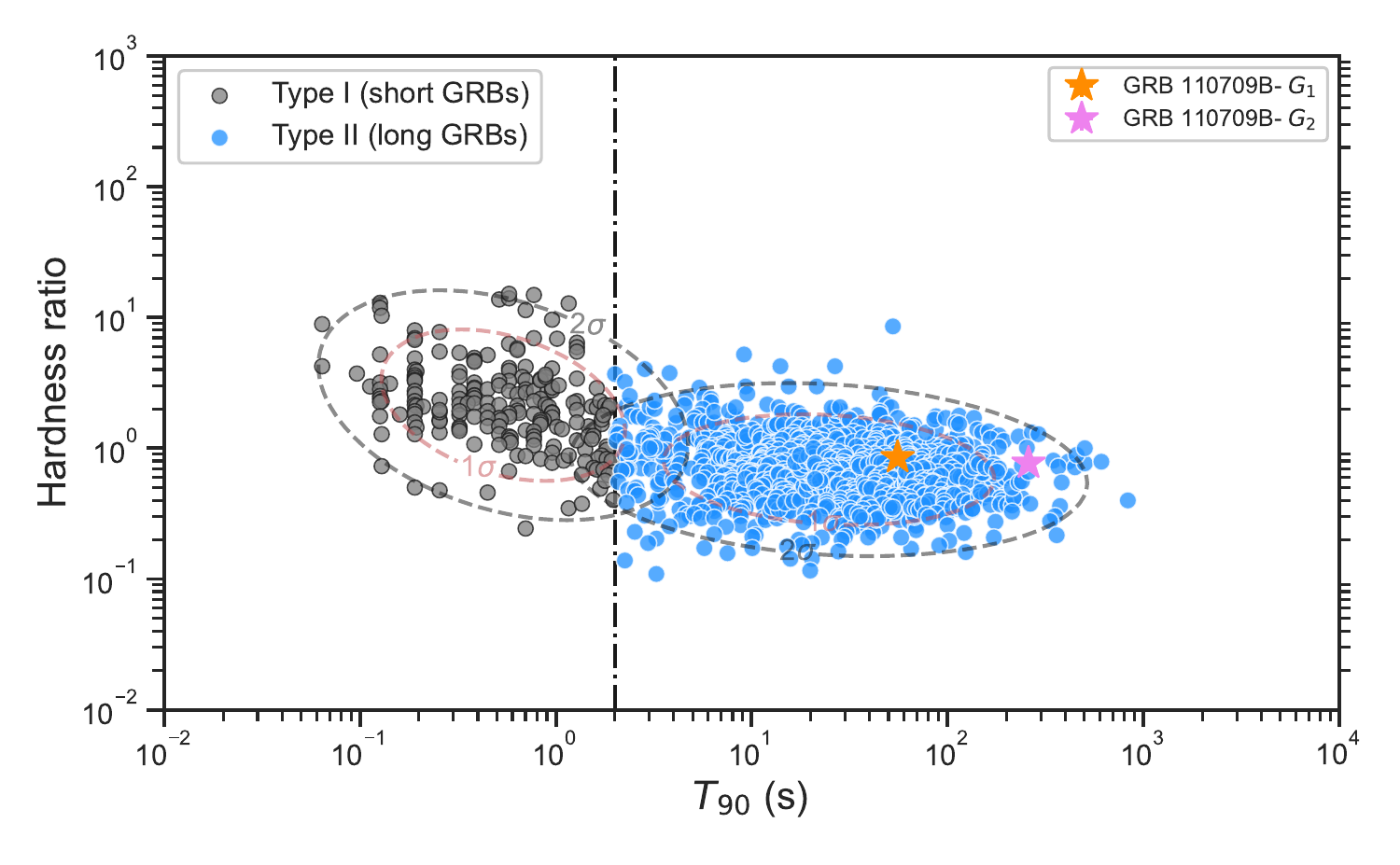}
{\bf b}\includegraphics[width=0.45\columnwidth]{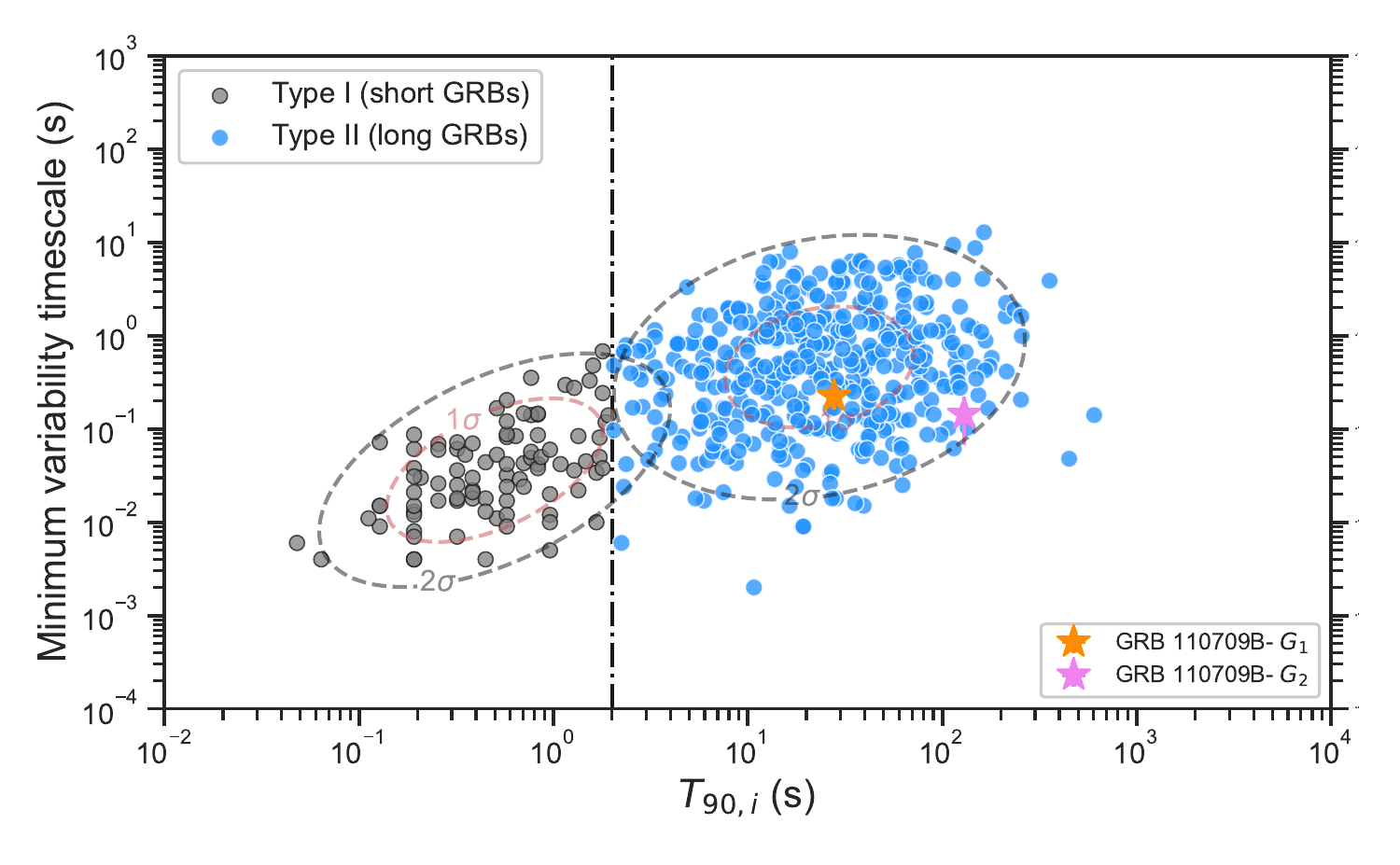}
\caption{Same as Figure \ref{fig:091024A_Classification} but for GRB 110709B.}
\label{fig:110709B_Classification}
\end{figure}

\clearpage

\clearpage
\begin{figure}[ht!]
{\bf a}\includegraphics[width=0.45\columnwidth]{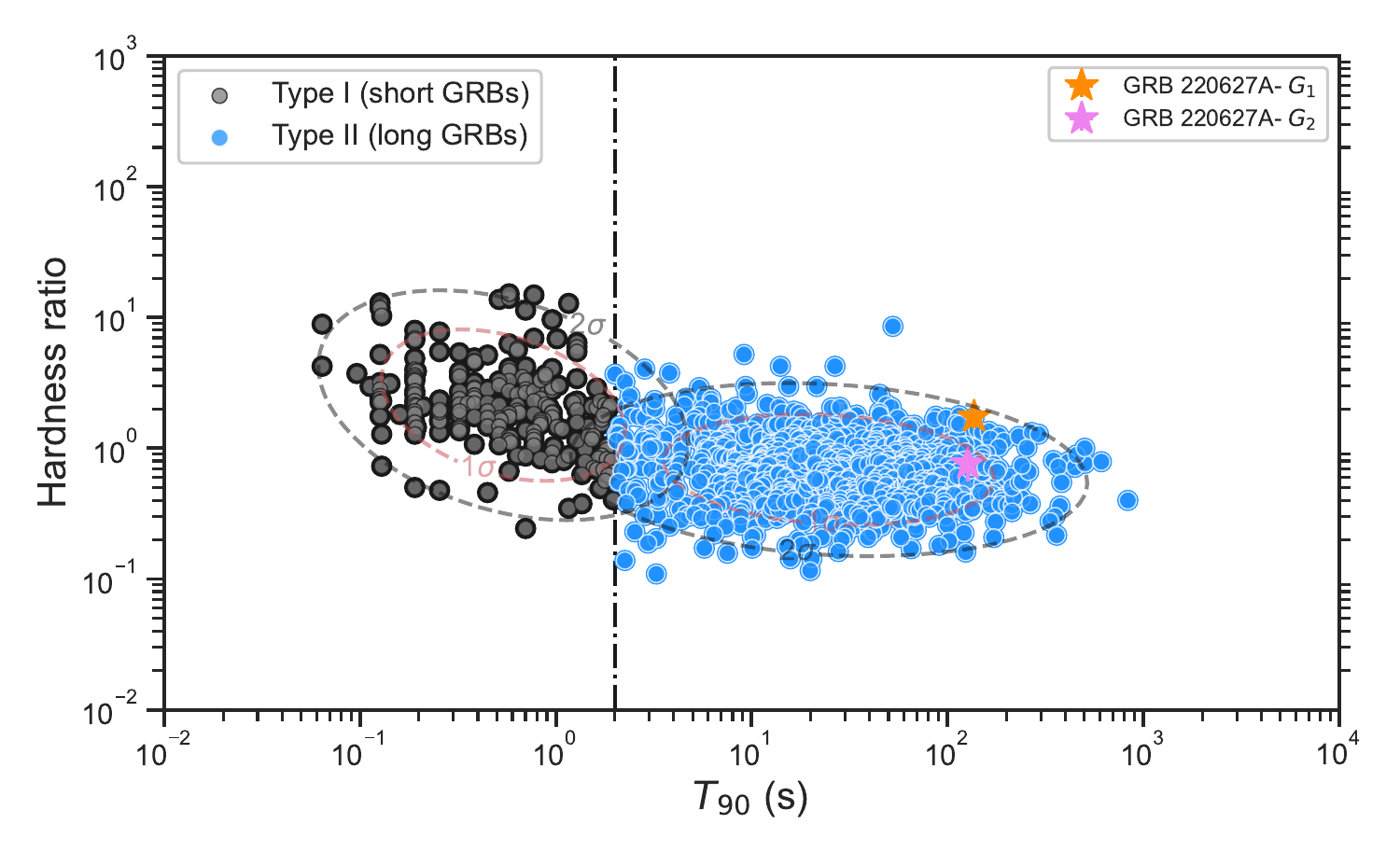}
{\bf b}\includegraphics[width=0.45\columnwidth]{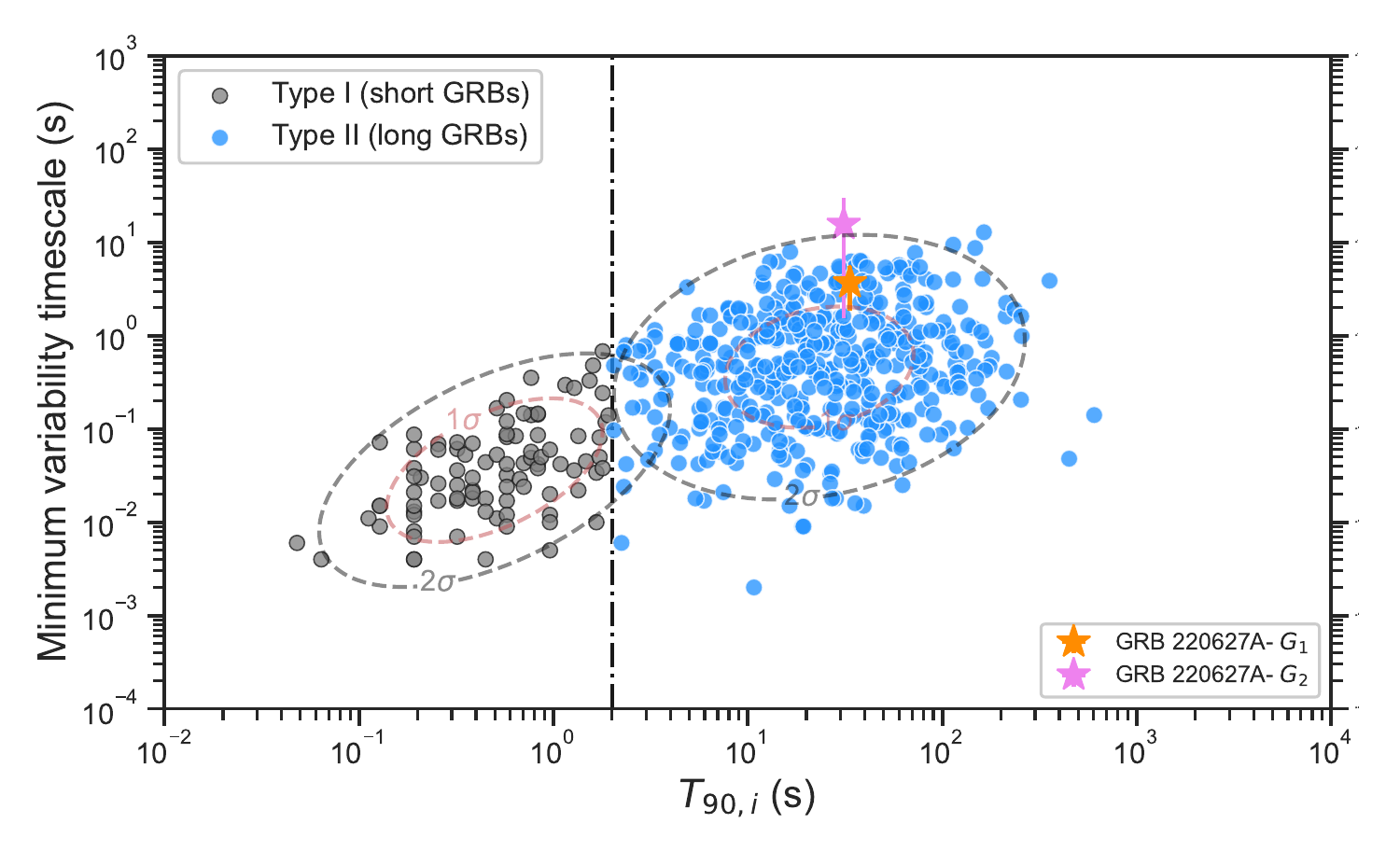}
{\bf c}\includegraphics[width=0.45\columnwidth]{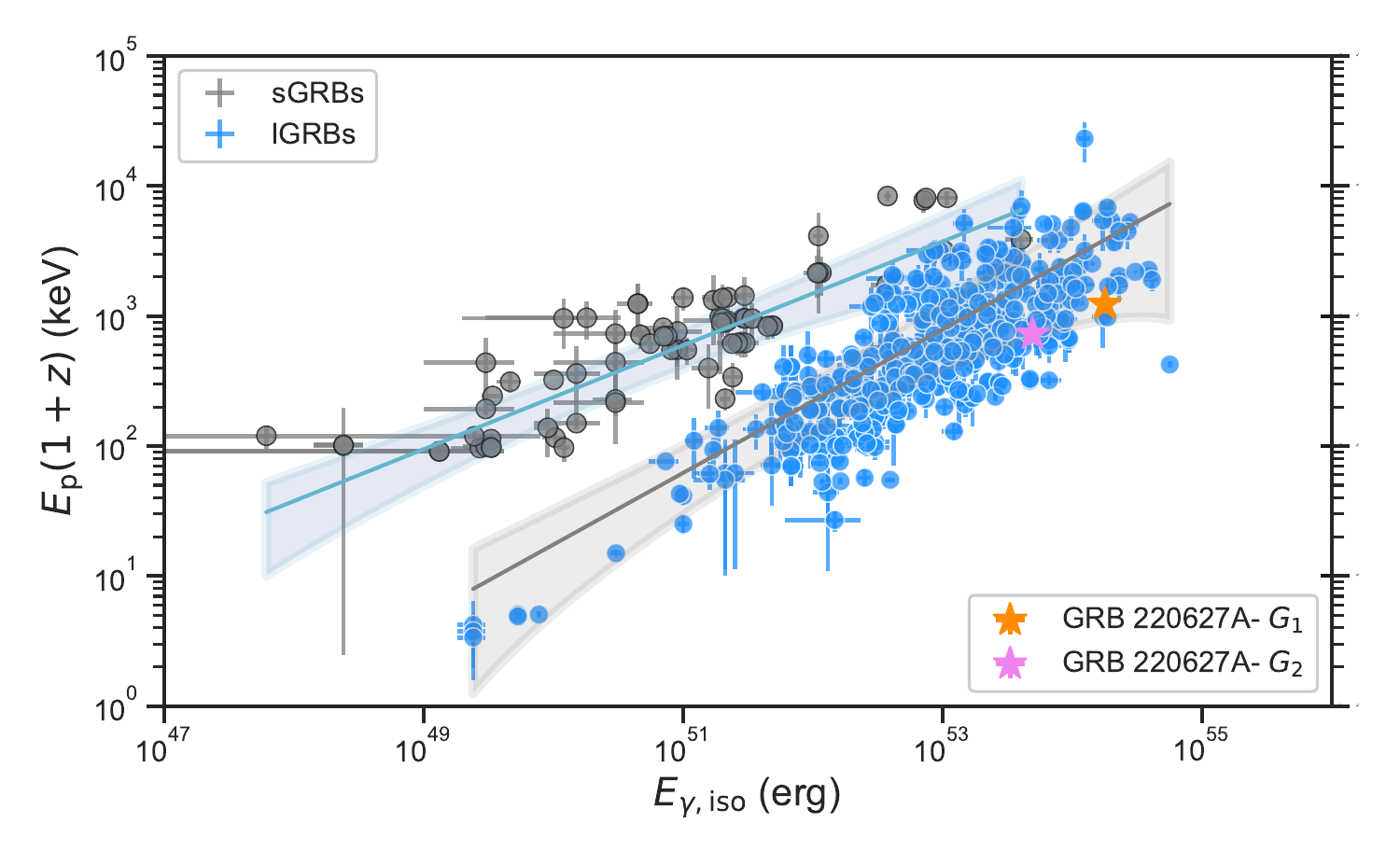}
{\bf d}\includegraphics[width=0.45\columnwidth]{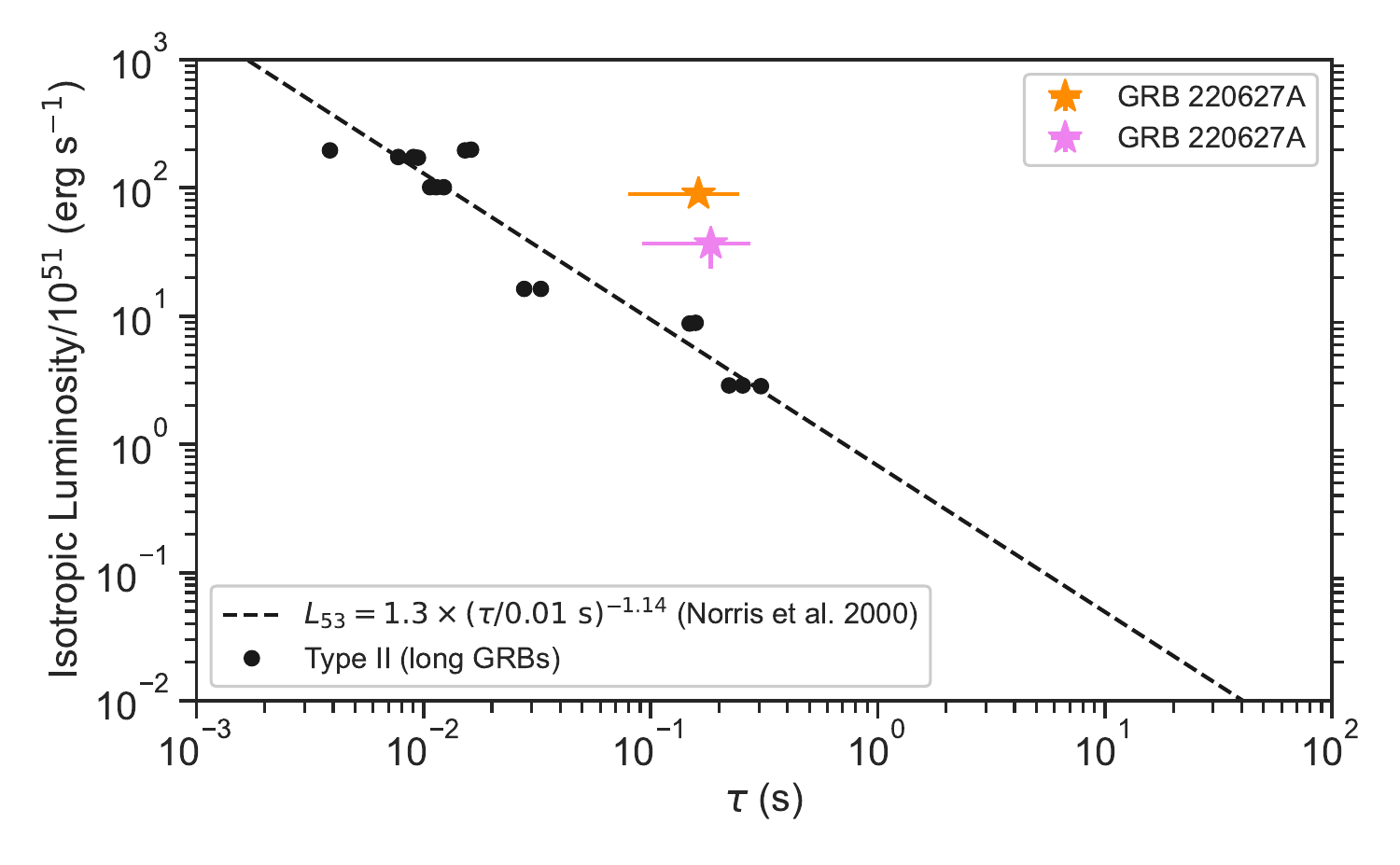}
{\bf e}\includegraphics[width=0.45\columnwidth]{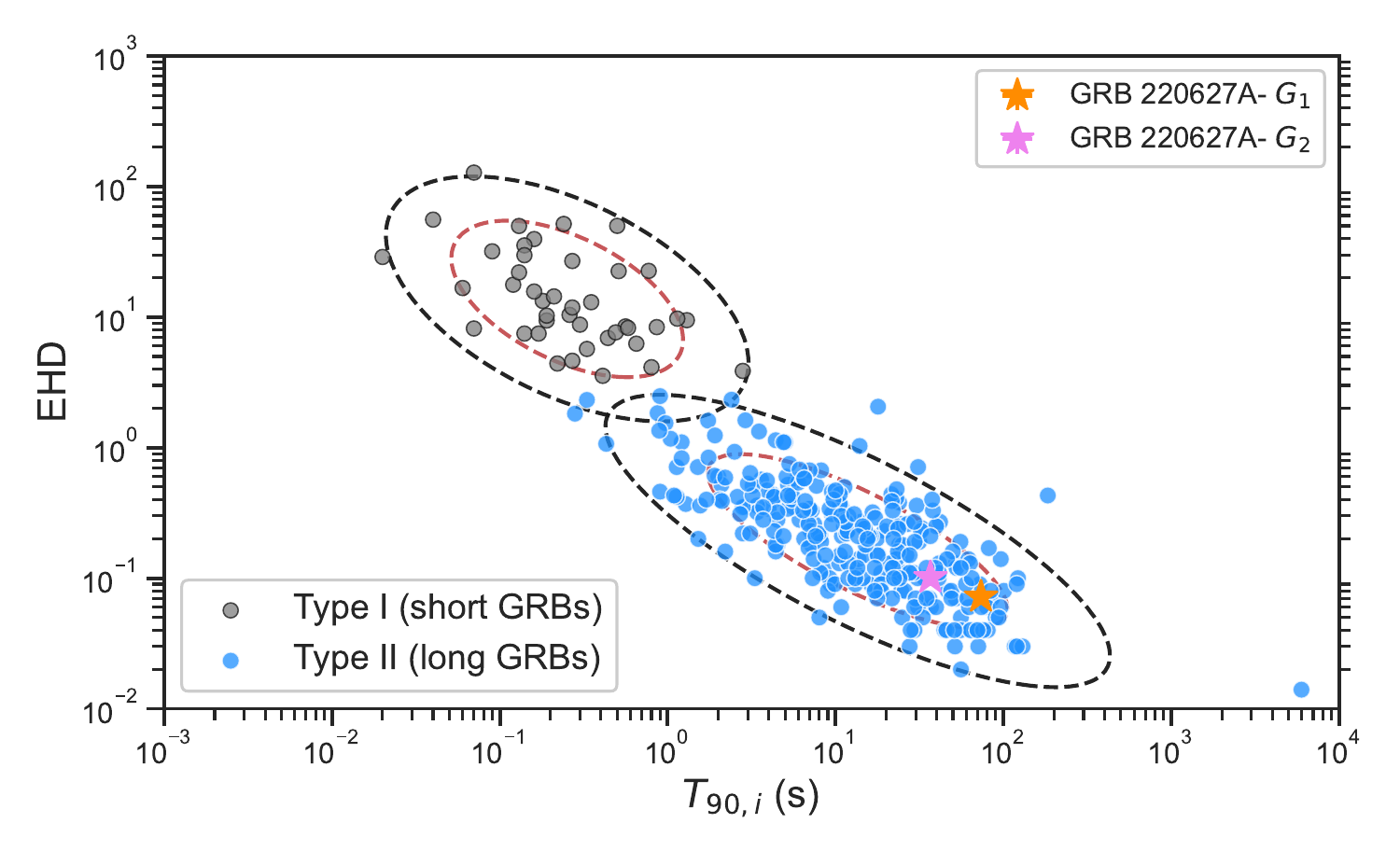}
{\bf f}\includegraphics[width=0.45\columnwidth]{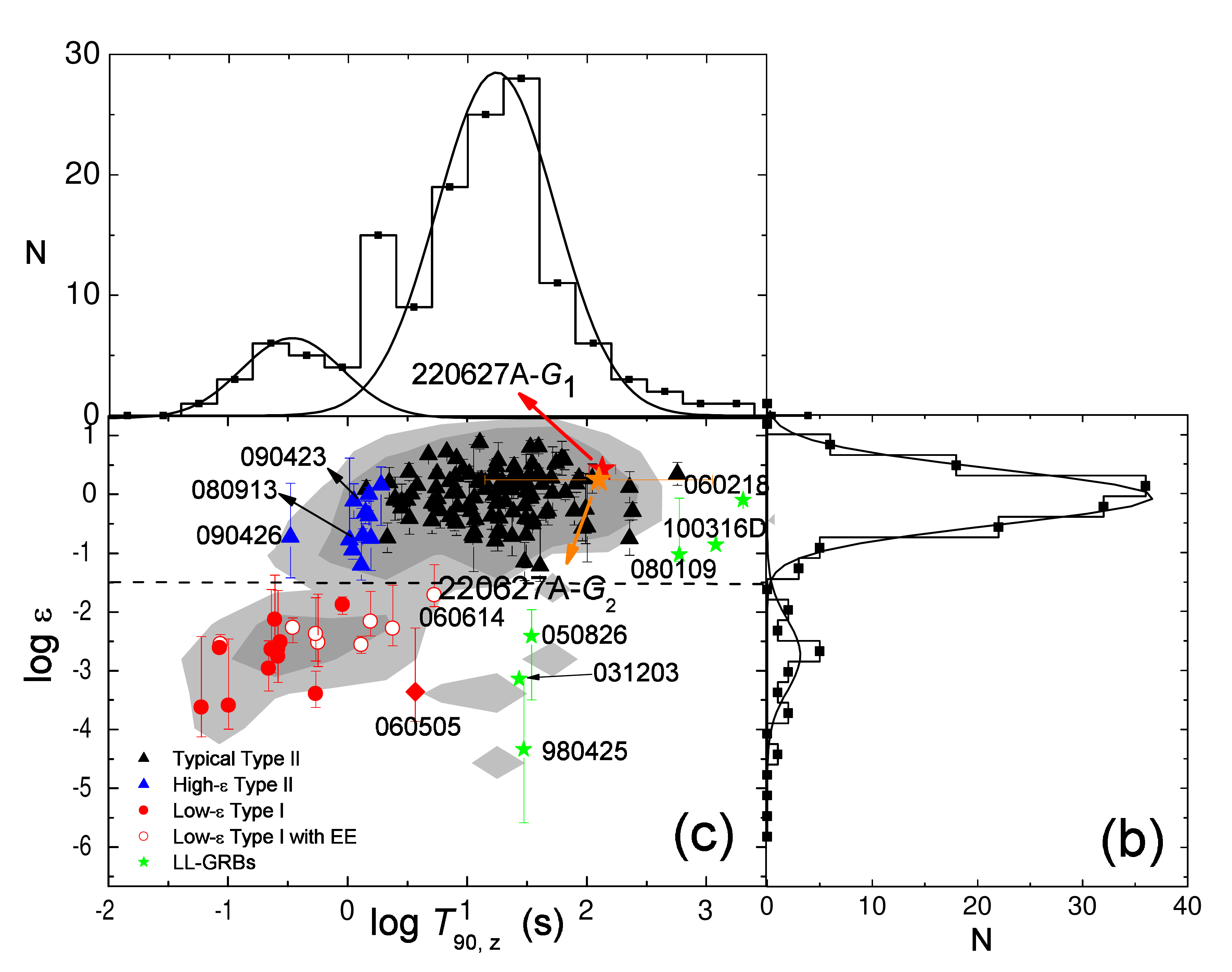}
\caption{Same as Figure \ref{fig:091024A_Classification} but for GRB 220627A. \label{fig:220627A_Classification}}
\end{figure}

\clearpage

\end{document}